\documentstyle[12pt]{article}
{\baselineskip 12pt}
\begin{document}
\begin{flushright}
{\bf WSP--IF 98--51\\
March 1998, \\ Revised: January 2000, \\
physics/9803030 }
\end{flushright}
\hfill \\
\begin{center}
{\Large \bf
Improved Lee, Oehme and  Yang  \\ \hfill \\
approximation. } \\
\hfill \\    {\em K. Urbanowski}$^{\ast}$
and {\em J. Piskorski }\\
Pedagogical University, Institute of Physics, \\
Plac Slowianski 6, 65-069 Zielona Gora, Poland. \\
\hfill \\
January 24, 2000.
\end{center}
\hfill \\
{\noindent}PACS numbers: 03.65.Bz., 11.10.St., 13.20.Eb.
\nopagebreak
\begin{abstract}
The Lee, Oehme and Yang (LOY) theory of  time evolution in two state   
subspace of states of the complete system is discussed. Some 
inconsistencies in assumptions and approximations used in  the standard
derivation of the LOY effective Hamiltonian, $H_{LOY}$, governig  this 
time evolution  are found. Eliminating these inconsistecies and 
using the LOY method, approximate  formulae for the effective 
Hamiltonian, $H_{\parallel}$,  governing the time evolution in this 
subspace (improving those obtained  by LOY) are derived. It is found, 
in  contradistinction  to  the  standard LOY result,  that in the 
case of neutral kaons $(<K_{0}|H_{\parallel}|K_{0}> - 
<\overline{K}_{0}|H_{\parallel}|\overline{K}_{0}>)$,  cannot take the 
zero value if  the  total system  the preserves CPT--symmetry. Within 
the use of the method mentioned above formulae for $H_{\parallel}$ 
acting in the three state (three dimensional) subspace of states are 
also found.
\end{abstract}
\hfill \\
\hfill \\
$^{\ast}$e--mail: kurban@omega.im.wsp.zgora.pl;
kurban@magda.iz.wsp.zgora.pl \\
\pagebreak[4]

\section[1]{Introduction.}
In the quantum decay theory of multiparticle complexes, properties    
of the transition amplitudes
\begin{equation}
A_{u_{j}; \psi }(t) = < u_{j}| \psi ;t>
\label{amp}
\end{equation}
are usually analysed. Here vectors $\{ |u_{j}> {\}}_{j \in U}$ 
represent the unstable states of the system considered, $<u_{j}|u_{k}> 
= {\delta}_{jk}$, and 
$| \psi ; t>$ is the solution of the Schr\"{o}dinger 
equation (we use $\hbar = c = 1$ units)
\begin{equation}
i \frac{\partial}{\partial t} |{\psi};t> = H 
|{\psi};t>,  \label{Sch}
\end{equation}
having the following form
\begin{equation}
| \psi ;t> = \sum_{j \in U}a_{j}(t) |u_{j}> + \sum_{J}
f_{J}(t)| {\phi}_{J}> , \label{psi-gen}
\end{equation}
where vectors $|{\phi}_{J}>$ describe the states of decay products, 
$<u_{j}|{\phi}_{J}> =0$ for every $j \in U$. The initial condition for 
Eq (\ref{Sch}) in the case considered is usually assumed to be 
\begin{eqnarray}
| \psi ;t = t_{0} \equiv 0 > & \stackrel{\rm def}{=} &
| \psi > \equiv \sum_{j \in U} a_{j} |u_{j}>, \label{init0} \\
f_{J} (t = 0) & = & 0. \nonumber
\end{eqnarray}
In Eq (\ref{Sch})  $H$ denotes the complete (full), selfadjoint 
Hamiltonian of the system. We have $| \psi ; t> =e^{-itH} | \psi >$. It 
is not difficult to see that this property and hermiticity of $H$ imply 
that
\begin{equation}
A_{u_{j},u_{j} }(t)^{\ast} =A_{u_{j};u_{j}}(-t) . \label{amp-ast}
\end{equation}
Therefore, the decay probability of an unstable state (usually called 
the decay law), i.e., the probability 
for a quantum system to remain in its initial state $| \psi > 
\equiv |u_{j}>$
\begin{equation}
p_{ u_{j} } (t) \stackrel{\rm def}{=} |A_{u_{j};u_{j}}(t)|^{2} \equiv 
|a_{j}(t)|^{2}, \label{prob}
\end{equation}
must be an even function of time:
\begin{equation}
p_{ u_{j} } (t) = p_{u_{j} } ( - t). \label{even}
\end{equation}

This last property suggests that in the case of the unstable states 
prepared at some instant $t_{0}$, say $t_{0} = 0$,  the initial 
condtion (\ref{init0}) for the evolution equation (\ref{Sch}) should 
be formulated more precisely. Namely, from (\ref{even}) it follows 
that the probabilities of finding the system in the decaying state 
$|u_{j}>$ at the instant, say $t=T \gg t_{0} \equiv 0$, and at the 
instant $t =-T$ are the same. Of course, this can never occur. In 
almost all experiments in which the decay law of a given unstable 
particle is investigated this particle is created at some instant of 
time, say $t_{0}$, and this instant of time is usually considered as 
the initial instant for the problem. From the property (\ref{even}) 
it follows that the instantaneous creation  of the 
unstable particle is impossible. For the observer, the 
creation of this particle (i.e., the preparation of the state, 
$|u_{j}>$, representing the decaying particle) is practically 
instantaneous. What is more, using suitable detectors he is usually 
able to prove that it did not exist at times $t < t_{0}$.  Therefore, 
if one looks for the solutions of the Schr\"{o}dinger equation 
(\ref{Sch}) describing properties of the unstable states prepared at 
some initial instant $t_{0}$ in the system, and if one requires these 
solutions to reflect situations  described above,
one should completement initial conditions (\ref{init0}) for Eq 
(\ref{Sch}) by assuming additionally that
\begin{equation}
a_{j}(t < t_{0}) = 0, \; \; ( j \in U), \label{init01}
\end{equation} 
and that, for the problem, time $t$ varies from $t = t_{0} > - 
\infty$ to $t = + \infty$ only.

Amplitudes of type $a_{j}(t)$ can be calculated directly by solving 
the evolution equation (\ref{Sch}),  or by using the 
Schr\"{o}dinger--like evolution equation governing the time evolution 
in a subspace spanned by the set of vectors $\{ |u_{j}> 
{\}}_{j \in U}$. Searching for the properties of two particle 
subsystems one usually uses the following equation of the type 
mentioned \cite{LOY1} --- \cite{tsai} instead of Eq (\ref{Sch}), 
\begin{equation}
i \frac{\partial}{\partial t} |\psi ; t >_{\parallel} =
H_{\parallel} |\psi ; t >_{\parallel},
\label{H-par}
\end{equation}
where by $H_{\parallel}$ we denote the effective nonhermitian 
Hamiltonian, 
\begin{equation}
H_{\parallel} \equiv M - \frac{i}{2} \Gamma, \label{H-par0}
\end{equation}
and
\begin{equation}
M = M^{+}, \; \; \Gamma = {\Gamma}^{+}, \label{M-G}
\end{equation}
are $(2 \times 2)$ matrices, acting in a two--dimensional subspace 
${\cal H}_{\parallel}$ of the total state space $\cal H$. $M$ is 
called the mass matrix, $\Gamma$ is the decay matrix \cite{LOY1} --- 
\cite{dafne}. The standard method of derivation of such a 
$H_{\parallel}$ is based on a modification of Weisskopf--Wigner (WW) 
approximation \cite{ww}. Lee, Oehme and Yang (LOY) 
adapted the WW aproach to the case of 
a two particle subsystem \cite{LOY1} --- \cite{5} to obtain their 
effective Hamiltonian $H_{\parallel} \equiv H_{LOY}$. Almost all 
properties of the neutral kaon complex, or another two state 
subsystem, can be described by solving Eq (\ref{H-par}) \cite{LOY1} 
--- \cite{tsai}, with the initial condition corresponding to 
(\ref{init0}) and (\ref{init01})
\begin{eqnarray}
| \psi ;t = t_{0} >_{\parallel} & \equiv &
| \psi >_{\parallel}, \nonumber \\
\parallel \, |{\psi} ; t = t_{0} >_{\parallel} 
{\parallel} & = & 1, \; \; | {\psi}; t < t_{0} >_{\parallel} = 0, 
\label{init}
\end{eqnarray}
for $| \psi ; t >_{\parallel}$ belonging to the subspace
${\cal H}_{\parallel} \subset {\cal H}$ spanned, e.g., by 
orthonormal neutral  kaons states $|K_{0}>, \; |{\overline{K}}_{0}>$, 
and so on, (then states corresponding to the decay products belong to 
${\cal H} \ominus{\cal H}_{\parallel} \stackrel{\rm def}{=} 
{\cal H}_{\perp}$),
\begin{equation}
|\psi >_{\parallel} \equiv a_{1}|{\bf 1}> + a_{2}|{\bf 2}>, 
\label{psi-par}
\end{equation}
and $|{\bf 1}>$ stands for  the  vectors of the   $|K_{0}>,  \; 
|B_{0}>$, etc., type and $|{\bf 2}>$ denotes states of
$|{\overline{K}}_{0}>, \; {\overline{B}}_{0}>$ type,
$<{\bf j}|{\bf k}> = {\delta}_{jk}$, $j,k =1,2$.  

The  old, as well as the more recent  \cite{4}
--- \cite{dafne} experimental   tests    of the  CP--noninvariance 
and of the CPT--invariance in the neutral  kaon  system need a correct  
interpretation of the measured CP-- nad CPT--violation parameters. 
In  the large literature, all CP--  and  CPT--violation
parameters in the neutral kaon and similar  complexes  are  expressed  
in terms  of  matrix elements  of  $H_{\parallel} \equiv H_{LOY}$.
On the other hand, in some papers the correctness and selfconsistency 
of the LOY approximation is questioned \cite{beyond} --- \cite{tsai}, 
\cite{kabir}. Therefore it seems to be important to examine in detail 
the derivation of the formulae for $H_{LOY}$. 

The paper is organized as follows. We begin with the discussion       
of the Lee, Oehme and Yang theory: Deriving formulae for matrix 
elements of $H_{LOY}$ in Sec. 2 we will apply the method used in 
\cite{Gaillard} with insignificant modifications. In Sec. 3 within the 
use of the same "recipe" as in Sec. 2, instead of the formulae 
for matrix elements, the formula for the complete operator $H_{LOY}$ 
is derived and the questionable points of the LOY approach are found. 
Namely in \cite{LOY1,LOY2} terms of type $<{\bf j}|H^{(1)}|{\bf k}>$, 
where $H^{(1)}$ denotes a small perturbation and $j,k =1,2$, 
are neglected in the initial equations for amplitudes $a_{j}(t)$. 
The aim
of this paper is to show that taking into account such terms with 
the use of all remaining LOY assumptions lead to the effective
Hamiltonian $H_{\parallel}$, which differs from $H_{LOY}$. 
This $H_{\parallel} = H_{LOY}^{Imp}$ improving 
$H_{LOY}$ is also found in this Section. The improved LOY method is 
used in Sec. 4 to derive the effective Hamiltonian $H_{\parallel}$ 
governing the time evolution in the three dimesional (three state) 
subspace of states. Sec. 5 contains a  summary and conclusions.

\section[2]{Analysis of steps leading to the standard formulae for 
$H_{LOY}$.}

\subsection{Detailed derivation of $H_{LOY}$.}

Let us now consider all the steps leading to the formulae for the 
matrix elements of  $H_{LOY}$ in detail. As it has already been 
mentioned, the source of the LOY model  for the  decay  of
neutral kaons is the well known Weisskopf--Wigner  approach  to  the 
description  of  unstable  states \cite{ww}.  Within  this   approach,   
the Hamiltonian $H$ for the problem is divided into two parts 
$H^{(0)}$ and $H^{(1)}$:  
\begin{equation}
H \; = \; H^{(0)} + H^{(1)},  \label{H}
\end{equation}
such that $|K_{0}> \equiv|{\bf 1}>$ and $| {\overline  K}_{0}> 
\equiv |{\bf 2}>$ are
discrete  eigenstates  of  $H^{(0)}$ for the 2--fold degenerate 
eigenvalue $m_{0}$, 
\begin{equation}
H^{(0)} |{\bf j} > = m_{0} |{\bf j }>, \; \;  j = 1,2 ; \label{Hm_0}
\end{equation}
and $H^{(1)}$ induces the transitions from  these  states  to
other  (unbound)  eigenstates  $|\varepsilon ,J >$  of  $H^{(0)}$  
(here $J$ denotes such quantum numbers as charge, spin, etc.),
and, consequently, also between $|K_{0}>$ and $| {\overline K}_{0} >$. 
So, the problem which one usually considers is the time evolution of 
an initial state, which is a superposition  of 
$|{\bf 1}>$ and  $|{\bf 2}>$ states.

In the kaon rest--frame, this time evolution  for $t \geq  t_{0} 
\equiv 0$ is  governed  by the Schr\"{o}dinger equation
(\ref{Sch}), whose solutions $|\psi ; t>$ have the following 
form 
\begin{equation}
|{\psi};t> = a_{1}(t)|{\bf 1}> + a_{2}(t)|{\bf 2}> +
\sum_{J, \varepsilon} F_{J}(\varepsilon ;t) |\varepsilon ,J>,
\label{psi}
\end{equation}
and
\begin{equation}
|a_{1}(t)|^{2} + |a_{2}(t)|^{2} + \sum_{J, \varepsilon} 
|F_{J}( \varepsilon ,t)|^{2} = 1. \label{psi-n}
\end{equation}
Here $|F_{J}; t> \equiv  \sum_{\varepsilon}  
F_{J}(\varepsilon ;t)| \varepsilon ,J >$ represents the decay 
products in the channel $J$;  $< \varepsilon ,J|{\bf k}> = 0$, 
$k = 1,2$; $< \varepsilon ',L| \varepsilon , N > = {\delta}_{LN} 
{\delta}(\varepsilon - \varepsilon ')$.  

From the Schr\"{o}dinger equation (\ref{Sch}) 
the  following equations   for amplitudes $a_{1}(t)$, $a_{2}(t)$ and 
$F_{J}(\varepsilon ;t)$ can be obtained 
\begin{eqnarray}
i \frac{\partial}{\partial t} a_{k}(t) &=& m_{0} a_{k}(t) +
\sum_{l = 1}^{2} H_{kl}^{(1)}a_{l}(t) \nonumber \\
&+&
\sum_{J, \varepsilon} H_{kJ}^{(1)}( \varepsilon ) F_{J}( \varepsilon 
;t), \; \; \; \; ({\scriptstyle k = 1,2}; \; \; t \geq 0)  
\label{a_k1} \\
i \frac{\partial}{\partial t} F_{J}( \varepsilon ;t) &=&
\varepsilon F_{J}( \varepsilon ;t) + \sum_{k=1,2}
H_{Jk}^{(1)}( \varepsilon ) a_{k}(t) \nonumber \\
&+&
\sum_{L, \varepsilon '} F_{L}( \varepsilon ';t) H_{J,L}^{(1)}( 
\varepsilon , \varepsilon '), \; \; \; \; (t \geq 0),
\label{F_j1}
\end{eqnarray}
where $H_{kJ}^{(1)}( \varepsilon ) = ( H_{Jk}^{(1)}( 
\varepsilon ))^{\ast} = <{\bf k}|H^{(1)}| \varepsilon ,J>$, 
($k =1,2$), are the matrix elements responsible for the decay, 
$H_{J,L}^{(1)}( \varepsilon ,\varepsilon ') = <J, \varepsilon 
|H^{(1)}| \varepsilon ',L>$,  and
$H_{kl}^{(1)}$ \linebreak  $= <{\bf k}|H^{(1)}|{\bf l}>$; $k,l = 1,2$.  
These equations are exact. In agreement with (\ref{init}), 
the boundary conditions for Eqs (\ref{a_k1}), 
(\ref{F_j1}) are following: 
\begin{equation}
a_{k}(0) = a_{k}, \; \; a_{k}(t<0) = 0, \; \; ({\scriptstyle k=1,2}), 
\label{a(0)}
\end{equation}
and 
\begin{equation}
F_{J}(\varepsilon ; t = 0) = 0,  \label{F(0)}
\end{equation}
so
\begin{equation}
|a_{1}|^{2} + |a_{2}|^{2} = 1.
\label{a(0)-1}
\end{equation}

In the WW approach to solving the Schr\"{o}dinger equation (\ref{Sch}) 
it is required that the martix elements of type $H_{jk}^{(1)}, 
H_{kJ}^{(1)}( \varepsilon )$, etc., should be suitably small \cite{ww}. 
From \cite{LOY1,LOY2} and \cite{Gaillard} one can conclude that
the LOY modification of the WW method consists of assuming that, among 
others,\begin{eqnarray}
\sum_{k=1,2}
|H_{jk}^{(1)}| &\ll & m_{0}, \; \; \; ({\scriptstyle j =1,2}),
\label{ww1a} \\
\sum_{J, \varepsilon}
|H_{kJ}^{(1)}( \varepsilon )| &\ll& m_{0},
\; \; \; ({\scriptstyle k =1,2}), \label{ww1b} \\
\sum_{l=1,2}
|H_{kl}^{(1)}| &\ll&  \sum_{J, \varepsilon}
|H_{kJ}^{(1)}( \varepsilon )|, 
\; \; \; ({\scriptstyle k =1,2}), \label{ww2}
\end{eqnarray}
and 
\begin{equation}
\sum_{L, \varepsilon '}
|H_{J,L}^{(1)}( \varepsilon ,\varepsilon ')| 
\ll  \sum_{k=1,2}
|H_{Jk}^{(1)}( \varepsilon )|,
\label{ww3} 
\end{equation}
for every $J$.

Assumptions of type (\ref{ww1a}) --- (\ref{ww3}) were used by LOY in 
order to replace the exact equations of type  (\ref{a_k1}), 
(\ref{F_j1}) by approximate equations (18) --- (20) considered in  
\cite{LOY1} (see also \cite{Gaillard}, Chap. 5, Appendix 1, Equations 
(A1.4) --- (A1.6)). The mentioned LOY equations are equivalent to the 
following approximate ones, which are valid if the requirements 
(\ref{ww1a}) --- (\ref{ww3}) hold 
\begin{eqnarray}
i \frac{\partial}{\partial t} a_{k}(t) &=& m_{0} a_{k}(t) 
+
\sum_{J,\varepsilon} H_{kJ}^{(1)}(\varepsilon ) F_{J}(\varepsilon ;t), 
\; \; \; \; ({\scriptstyle k = 1,2}),  \label{a_k2} \\
i \frac{\partial}{\partial t} F_{J}( \varepsilon ;t) &=&
\varepsilon F_{J}( \varepsilon ;t) + \sum_{k=1,2}
H_{Jk}^{(1)}( \varepsilon ) a_{k}(t). 
\label{F_j2}
\end{eqnarray}

Eqs (\ref{a_k2}), (\ref{F_j2}) differ from LOY Eqs (18) --- (20) of 
\cite{LOY1}, among others, in the first  componets of 
their right sides . Such 
components are absent in the LOY equations. This difference is 
caused by using the interaction representation in \cite{LOY1} 
and rescaling the energy, $\varepsilon$: $\varepsilon 
\rightarrow\omega  =  \varepsilon  -  m_{0}$, which means that the 
zero of energy is taken to be the rest energy  of $K$. Another 
difference is the following: In the right sides of the LOY equations 
factors of type $e^{\textstyle \pm i \omega t}$ are present. They are 
absent in Eqs (\ref{a_k2}), (\ref{F_j2}). The presence of these 
factors  in LOY equations is due  to the use of the interaction 
representation. Nevertheless, the 
mathematical equivalence of Eqs (\ref{a_k2}), (\ref{F_j2}) and  Eqs 
(18) --- (20) of \cite{LOY1} is rigorous.

The WW theory states that under the assumptions (\ref{ww1a}) --- 
(\ref{ww3}), the actual contribution  of the second component on the 
rigth side of Eq (\ref{a_k2}) into the amplitude $a_{k}(t)$ is very 
small. From \cite{LOY1,LOY2,Gaillard} one can conclude that in 
the LOY treatment of the problem this contribution 
resolves itself into adding some small complex 
number, say $\Lambda$, to the parameter $m_{0}$, such that $| \Lambda 
| \ll m_{0}$, and $ {\rm Im.} \, \Lambda =- \frac{\gamma}{2} < 0$. 
Simply, the interactions which are responsible for the presence of 
this second component in the considered equation  slightly shift the 
level $m_{0}$: $m_{0} \rightarrow m_{0} + \Lambda$. So, the 
replacement of Eq (\ref{a_k2}) by the following approximate one seems 
to be justifiable
\begin{equation}i \frac{\partial}{\partial t} a_{k}(t)  \simeq (m_{0} 
+ \Lambda )a_{k}(t),
\; \; \; \; ({\scriptstyle k = 1,2}; \; t > 0).  \label{a_k2a}
\end{equation}
which means that under the conditions (\ref{ww1a}) --- (\ref{ww3}), 
the amplitudes $a_{k}(t)$ should take the following form
\begin{equation}
a_{k}(t) \simeq e^{-i(m_{0} + \Lambda )t} a_{k}, \; \; 
({\scriptstyle k=1,2}; \; t > 0),
\label{a_k2a-ex}
\end{equation}
Therefore, when one looks for the solutions of Eq (\ref{a_k2}), the 
use of the assumption 
\begin{equation}
\frac{a_{1}(t)}{a_{1}} = \frac{a_{2}(t)}{a_{2}} = 
e^{-i(m_{0} + \Lambda )t},  \; \; (t > 0),
\label{main-as}
\end{equation}
is considerd to be obvious. This assumption is equivalent to the 
LOY assumption (21) of \cite{LOY1} (or, (A1.1) in \cite{Gaillard}, 
Appendix of Chap. 5), which is easily seen if (\ref{main-as}) is 
rewritten in the LOY manner:
\begin{equation}
| \psi ; t >_{\parallel} = e^{-i(m_{0} + \Lambda )t} | \psi 
>_{\parallel}, \; \; (t > 0),
\label{LOY-as}
\end{equation}
where
\begin{equation}
| \psi ; t >_{\parallel} = a_{1}(t)|{\bf 1}> + a_{2}(t)|{\bf 2}>.
\label{psi-par-t}
\end{equation}  
The assumption (\ref{main-as}) (or (\ref{LOY-as}) ) is 
crucial to the LOY method and it is the essence of the approximation  
which was made in \cite{LOY1,Gaillard}. It determines all the 
properties  of the  effective  Hamiltonian $H_{LOY}$ governing the time  
evolution  in a  two state  subspace.

Defining
\begin{equation}
F_{J}( \varepsilon ;t) \stackrel{\rm def}{=} e^{-i \varepsilon t}
{\tilde{F}}_{J}( \varepsilon ;t), \label{F-tilde}
\end{equation}
Eq (\ref{F_j2}) can be transformed into
\begin{eqnarray}
i \frac{\partial}{\partial t} {\tilde{F}}_{J}( \varepsilon ;t) &=&
\sum_{k=1,2}
e^{i \varepsilon t}
H_{Jk}^{(1)}( \varepsilon ) a_{k}(t) ,
\label{F_j3} \\
{\tilde{F}}_{J}( \varepsilon ; t = 0) & = & 0, \nonumber
\end{eqnarray}
which can easily be solved and leads to the following solution for
$F_{J}( \varepsilon ;t)$ with $t \geq 0$:
\begin{equation}
F_{J}( \varepsilon ;t) = ( -i ) \sum_{k=1,2}
\int_{0}^{t} e^{-i \varepsilon (t - \tau ) }
H_{Jk}^{(1)}( \varepsilon ) a_{k}( \tau ) d \tau .
\label{F_j4}
\end{equation}

Now, one can eliminate $F_{J}( \varepsilon ;t)$ from Eq (\ref{a_k2}) 
by substituting  (\ref{F_j4}) back into Eq (\ref{a_k2}). This leads 
to the following equation, eg., for $a_{1}(t)$ with $t \geq 0$, 
\begin{equation}
i \frac{\partial}{\partial t} a_{1}(t) = m_{0} a_{1}(t) 
-i \, \sum_{k=1,2} \, 
\sum_{J, \varepsilon} \, 
\int_{0}^{t} e^{-i \varepsilon (t - \tau ) }
H_{1J}^{(1)}( \varepsilon )
H_{Jk}^{(1)}( \varepsilon ) a_{k}( \tau ) \, d \tau . 
\label{a_k3}  
\end{equation}
Next, inserting (\ref{main-as}) into (\ref{a_k3}) one finds the
following equation for   $a_{1}(t)|_{t > 0}$  
\begin{eqnarray}
\Big\{
i \frac{\partial}{\partial t}  - m_{0} \Big\} a_{1}(t) 
&=& (-i) \sum_{k=1,2} \Big\{
\sum_{J, \varepsilon} 
\int_{0}^{t} e^{-i( \varepsilon  - m_{0} - \Lambda )(t - \tau ) }
H_{1J}^{(1)}( \varepsilon ) \times \nonumber \\
& &  \hspace{1.0in} 
\times 
H_{Jk}^{(1)}( \varepsilon ) d \tau  \; \Big\} 
a_{k}(t) . 
\label{a_k4}    
\end{eqnarray}

The main properties of the quasistationary, or bound states manifest 
themselves at times $t \gg t_{0} = 0$, where $t_{0}$ is the moment of 
their preparation. Therefore, it is reasonable to replace the upper 
limit $t < \infty$ of the integrals in Eq (\ref{a_k4}) by $t 
\rightarrow \infty$. Also, as it was mentioned, $\Lambda$ is a very 
small number. So, the formulae for 
the lowest notrivial order of the matrix elements $h_{jk}^{LOY}$ of 
$H_{LOY}$, are obtained by putting $\Lambda = 0$ under the integrals 
in Eq (\ref{a_k4}) and  then evaluating these integrals and passing 
to the limit  $t \rightarrow \infty$. (In this case these matrix 
elements will be denoted by $h_{jk}^{LOY(0)}$, and the the effective 
Hamiltonian by $H_{LOY}^{(0)}$). Such a treatment of Eq (\ref{a_k4}) 
gives (compare \cite{Gaillard} ) 
\begin{eqnarray}
\Big\{i \frac{\partial}{\partial t}  - m_{0} \Big\}a_{1}(t) 
& \simeq & -  \sum_{k=1,2} \, 
\Big\{  \lim_{t \rightarrow \infty}
\sum_{J, \varepsilon} 
\frac{1 - e^{-i( \varepsilon - m_{0})t}}{\varepsilon - m_{0}}
\times \nonumber \\
& & \hspace{.5in} \times
H_{1J}^{(1)}( \varepsilon )
H_{Jk}^{(1)}( \varepsilon ) \Big\} \,  a_{k}(t). \label{a_k5} 
\end{eqnarray}
where $t \gg t_{0} =0$. This last equation can be rewritten  as 
follows
\begin{equation}
\Big\{
i \frac{\partial}{\partial t}  - m_{0} \Big\} a_{1}(t)
= - {\Sigma}_{11}^{(0)}(m_{0}) a_{1}(t) 
- {\Sigma}_{12}^{(0)}(m_{0}) a_{2}(t), 
\label{a_k6}
\end{equation}
where $t \gg t_{0} = 0$, and 
\begin{equation}
{\Sigma}_{jk}^{(0)}(x) = 
\sum_{J, \varepsilon} 
H_{jJ}^{(1)}( \varepsilon ) \frac{1}{ \varepsilon - x - i0}
H_{Jk}^{(1)}( \varepsilon ) = <{\bf j}| {\Sigma}^{(0)}(x)|{\bf k}> .
\; \; ({\scriptstyle j,k =1,2} ).
\label{Sigma0-jk}
\end{equation}
A similar equation can be obtained for the amplitude $a_{2}(t)$. This 
means that the matrix elements $h_{jk}^{LOY(0)} = <{\bf j}
|H_{LOY}^{(0)}|{\bf k}>$  equal 
\begin{equation}
h_{jk}^{LOY(0)} = m_{0} {\delta}_{jk} - {\Sigma}_{jk}^{(0)}(m_{0}) 
\equiv M_{jk} - \frac{i}{2} {\Gamma}_{jk} , \; \; 
({\scriptstyle j,k = 1,2}), \label{h-LOY-jk}
\end{equation}
i.e., exactly as in \cite{LOY1} --- \cite{leonid}. 

These formulae are the frame for almost all
calculations  of the parameters characterizing the properties        
of the neutral  kaons complex  and  other two level subsystems 
\cite{leonid,baldo}.

\subsection{Operator form of $H_{LOY}$.}

Defining projectors
\begin{eqnarray}
P & = & |{\bf 1}><{\bf 1}| + |{\bf 2}><{\bf 2}|, \label{P}  \\
Q & = & I - P, \label{Q}
\end{eqnarray}
\begin{equation}
[P, H^{(0)}] = 0, \; \; \; \; [P, H^{(1)}] \neq 0, \label{com}
\end{equation}
(where $I$ is the unit operator in $\cal H$), 
allows us to rewrite 
the  $H_{LOY}^{(0)}$ in  a compact form which is
sometime more convenient than the standard one (\ref{h-LOY-jk}):
\begin{equation}
H_{LOY}^{(0)} = m_{0} P - {\Sigma}^{(0)}(m_{0})
\equiv M^{LOY} - \frac{i}{2} {\Gamma}^{LOY} ,
\label{H-LOY}
\end{equation}
where
\begin{equation}
{\Sigma}^{(0)} (x )  = PHQ \frac{1}{ H^{(0)} - x - i0 }
QHP. 
\label{Sigma0}
\end{equation}

The $H_{LOY}^{(0)}$ acts in a two dimensional subspace 
${\cal H}_{\parallel}$ of $\cal H$. This ${\cal H}_{\parallel}$ can be 
defined by  means of the projector $P$ in the following way 
\begin{equation}
{\cal H}_{\parallel}
\stackrel{\rm def}{=} P{\cal H} \ni | \psi ;t>_{\parallel},
\label{Hu}
\end{equation}
where
\begin{equation}
|\psi ;t>_{\parallel} \equiv P |\psi ;t> .
\label{psi-P}  
\end{equation}
The projector $Q$ defines the subspace  of  decay products 
${\cal H}_{\perp}$:
\begin{equation}
{\cal H}_{\perp} \stackrel{\rm def}{=} Q{\cal H}
\equiv {\cal H} \ominus {\cal H}_{\parallel} \ni | \psi ; t>_{\perp} ,
|F_{J};t>,
\label{Hd}
\end{equation}
where
\begin{equation}
|\psi ;t>_{\perp} \stackrel{\rm def}{=} Q |\psi ;t>. 
\label{psi-Q}  
\end{equation} 

Note that assumptions used in \cite{LOY1,Gaillard} lead to 
following the property
\[ 
\sum_{j = 1,2} |{\bf j}><{\bf j}| +
\sum_{J, \varepsilon} | \varepsilon ,J ><J,  \varepsilon |
= I.
\]
This means that the standard LOY approach enable us to conclude
that 
\[
I - P \equiv 
Q \equiv \sum_{J, \varepsilon} | \varepsilon ,J ><J,  \varepsilon | . 
\]
One should stress it that in a general case this last relation need 
not be valid and it will not be used in subsequent Sections of 
this paper.

\subsection{CPT transformation properties of $H_{LOY}$.}

Usually, in the LOY and related approaches, it is assumed that the 
free Hamiltonian $H^{(0)}$ is CPT--invariant \cite{LOY1} --- 
\cite{dafne}:
\begin{equation}
[{\Theta} , H^{(0)}] = 0 , 
\label{cpt-H0}
\end{equation}
where $\Theta$ is the antiunitary operator:
\begin{equation}
\Theta \stackrel{\rm def}{=} {\cal C}{\cal P}{\cal T},
\label{cpt}
\end{equation}
and  $\cal C $ is the charge conjugation operator, $\cal P$ --- space 
inversion, and the antiunitary operator $\cal T$ represents the time 
reversal operation. (Basic properties of anti--linear and linear 
operators, their products and commutators are described, eg., in 
\cite{cpt,messiah,bohm}). 

Using, e.g., the following phase convention \cite{LOY2} --- \cite{5}
\begin{equation}
\Theta |{\bf 1}> \stackrel{\rm def}{=} - |{\bf 2}>, \;\; 
\Theta|{\bf 2}> 
\stackrel{\rm def}{=} - |{\bf 1}>, \label{cpt1}
\end{equation}
which means that the subspace of neutral kaons ${\cal H}_{\parallel}$ 
is assumed to be invariant under $\Theta$:
\begin{equation}
[{\Theta}, P ] = 0, \label{cpt-P}
\end{equation}
one easily finds from  (\ref{h-LOY-jk}) that in the case of the      
CPT--invariant interactions 
\begin{equation}
[{\Theta} ,H^{(1)}] = 0, \label{cpt-H1}
\end{equation}
i.e., in the CPT--invariant system
\begin{equation}
[ \Theta , H] = 0, \label{cpt-H}
\end{equation}
the diagonal matrix elements of $H_{LOY}^{(0)}$ must be equal:  
\begin{equation}
h_{11}^{LOY(0)} = h_{22}^{LOY(0)}.  \label{h11=h22}
\end{equation}
This  is  the standard result of the LOY 
approach and this is the picture  which one meets in  the  literature  
\cite{LOY1}  ---  \cite{chiu}.

\section{Improved LOY approximation.}
\subsection{General considerations.}

In the previous Section the coupled system  equations (\ref{a_k1}), 
(\ref{F_j1}) for number functions (amplitudes) $a_{k}(t)$, 
$F_{J}( \varepsilon ;t)$ have been analysed. While considering each 
of the equations separately there is a danger of overlooking some 
common, global properties of a such system and thus similar properties 
of the physical system under consideration. It seems that a complex 
look at the equations governig the time evolution in the subsystem 
considered should either confirm all the conclusions and formulae 
derived above or show that they are incorrect. It 
should also indicate  all the questionable steps in the 
standard derivation of $H_{LOY}$. So, 
let us consider the evolution equations for the 
components $| \psi ;t>_{\parallel}$ (\ref{psi-par}), (\ref{psi-P})  
and  for $| \psi ; t>_{\perp}$ (\ref{psi-Q}) of the state vector 
$| \psi ;t>$ (\ref{psi}) instead of the system equations for number 
functions $a_{k}(t)$, $F_{J}( \varepsilon ;t)$. Using projection 
operators $P$ and $Q$, (\ref{P}), (\ref{Q}), one can obtain from 
the Schr\"{o}dinger equation (\ref{Sch}) for the state vector 
$| \psi ;t>$ two equations for its 
orthogonal components $| \psi ;t>_{\parallel}$ (\ref{psi-par}), 
(\ref{psi-P})  and $| \psi ; t>_{\perp}$ (\ref{psi-Q}) valid for $t 
\geq t_{0} = 0$:
\begin{eqnarray}
i \frac{\partial}{\partial t} | \psi ; t>_{\parallel} & = &
PHP | \psi ;t>_{\parallel} + PHQ | \psi ; t>_{\perp}, 
\label{eq-psi-par1} \\
& \equiv & \Big\{ m_{0}P + PH^{(1)}P \Big\} 
| \psi ;t>_{\parallel} + PH^{(1)}Q | \psi ; t>_{\perp}, 
\label{eq-psi-par2} \\
i \frac{\partial}{\partial t} | \psi ; t>_{\perp} & = &
QHQ | \psi ;t>_{\perp} + QHP | \psi ; t>_{\parallel}, 
\label{eq-psi-perp1} \\
& \equiv & QHQ | \psi ;t>_{\perp} + QH^{(1)}P | \psi ; t>_{\parallel}, 
\label{eq-psi-perp2}
\end{eqnarray}
with the initial conditions ({\ref{init}), (\ref{psi-par}) and 
(\ref{F(0)}), which are equivalent to the following one
\begin{equation}
| \psi ;t = 0>_{\perp} = 0. \label{psi-perp(0)}
\end{equation}

Let us consider a general case of Eqs (\ref{eq-psi-par1}) and 
(\ref{eq-psi-perp1}). According to  the LOY method, as in the usual 
single line width problem of atomic transitions \cite{ww}, the 
contribution arising from decay products $| \psi ;t>_{\perp} \in 
{\cal H}_{\perp}$ into the time 
derivative $i \frac{\partial}{\partial t} | \psi ;t>_{\parallel}$  
in Eq (\ref{eq-psi-par1}) should be eliminated.  Within this method, 
assuming that such a contribution is suitably small, one requires 
$i \frac{\partial}{\partial t} | \psi ;t>_{\parallel}$ to be expressed 
in terms of $| \psi ;t>_{\parallel}$ only. 
From the superposition principle Lee and Wu conclude in \cite{LOY2}
that  such an expression should be time independent and linear. 
Using this observation we find that
to fulfill this 
requirement, if the transitions from the subspace of decay products 
${\cal H}_{\perp} \ni | \psi ;t>_{\perp}$ are sufficiently weak (see 
(\cite{ur1})), i.e., if for every finite $t \geq 0$,
\begin{equation}
\parallel PHQ
| \psi ;t>_{\perp} \parallel \; \; \ll \; \; 
\parallel  PHP |\psi ;t>_{ \parallel} \parallel , \label{weak1}
\end{equation}
the following substitution into Eq (\ref{eq-psi-par1})  should be made
\begin{equation}
PHQ | \psi ; t>_{\perp}  = PH^{(1)}Q | \psi ; t>_{\perp} 
\equiv V_{\parallel} | \psi ;t>_{\parallel},
\label{V-par}
\end{equation}
where $V_{\parallel}$ is in general an linear and nonhermitian 
operator (a nonhermitian matrix) acting in the subspace 
${\cal H}_{\parallel} \ni | \psi ;t>_{\parallel}$. It is
additionaly assumed in \cite{LOY2} that an operator of this type 
should be time independent.
Then, to a very good approximation, 
Eqs (\ref{eq-psi-par1}), (\ref{eq-psi-par2}) 
take the required form 
\begin{eqnarray}
i \frac{\partial}{\partial t} | \psi ; t>_{\parallel} & = &
\Big\{ PHP  +  V_{\parallel} \Big\}  | \psi ; t>_{\parallel} 
\label{eq-psi-par3} \\
& \equiv & \Big\{ m_{0}P + PH^{(1)}P + V_{\parallel}  \Big\} 
| \psi ;t>_{\parallel}.  
\label{eq-psi-par4} 
\end{eqnarray}
This (within the use of the LOY assumption of time independence 
$V_{\parallel}$) means that  
one should expect the solutions of 
(\ref{eq-psi-par1}), (\ref{eq-psi-par2}) to have the exponential, 
similar to (\ref{main-as}) and (\ref{LOY-as}), form:
\begin{eqnarray}
| \psi ;t>_{\parallel} & = & 
e^{\textstyle -it(PHP + V_{\parallel})} | \psi ; t=0>_{\parallel} 
\label{ex-psi1}  \\
& \equiv &
e^{\textstyle -it( m_{0}P + PH^{(1)}P + V_{\parallel})} 
| \psi ; t=0>_{\parallel},
\label{ex-psi2}
\end{eqnarray}
and, as it has been done in the LOY theory,  such a form of 
$| \psi ;t>_{\parallel}$ can be used for the calculation of the 
effective Hamiltonian $H_{\parallel}$,
\begin{eqnarray}
H_{\parallel} &\stackrel{\rm def}{=}& PHP + V_{\parallel} 
\label{H-par-def1} \\
& \equiv & m_{0} P + PH^{(1)}P + V_{\parallel},  \label{H-par-def2}
\end{eqnarray}
governing the time evolution in the subspace considered.
 
Solving Eq (\ref{eq-psi-perp1}) one can eliminate $| \psi ;t>_{\perp}$ 
from Eq (\ref{eq-psi-par1}) by substituting the solution of Eq 
(\ref{eq-psi-perp1}) back into Eq (\ref{eq-psi-par1}). Looking for this 
solution we will follow the method used to solve Eq (\ref{F_j1}) in 
Sec. 2. Namely, by means of the substitution
\begin{equation}
|\widetilde{ \psi ;t} >_{\perp}  \stackrel{\rm def}{=} 
e^{\textstyle +itQHQ} | \psi ; t>_{\perp},  \; \; (t \geq 0),
\label{psi-tilde}
\end{equation}
Eq (\ref{eq-psi-perp1}) can be replaced by the following one
\begin{eqnarray}
i \frac{\partial}{\partial t}
|\widetilde{ \psi ;t } >_{\perp}  & = &
e^{\textstyle +itQHQ} QHP | \psi ; t>_{\parallel},  \; \; (t \geq 0),
\label{eq-psi-tilde} \\
| \widetilde{\psi ;t} = 0>_{\perp}  & = & 0. \nonumber
\end{eqnarray}
It is easy to solve this equation. Using its solution one finds
\begin{equation}
| \psi ;t>_{\perp} = -i \int_{0}^{t}
e^{\textstyle -i (t - \tau ) QHQ } QHP | \psi ; \tau >_{\parallel}
\, d \tau ,  \; \; (t \geq 0), \label{psi-perp1}
\end{equation}
which is in perfect agreement with the result (\ref{F_j4}) in Sec. 2.

Substituting (\ref{psi-perp1}) back into Eq (\ref{eq-psi-par1}) gives
for $t \geq 0$:
\begin{equation}
i \frac{\partial}{\partial t} | \psi ; t>_{\parallel}  = 
PHP | \psi ;t>_{\parallel} 
-i \int_{0}^{t} PHQ
e^{\textstyle -i (t - \tau ) QHQ } QHP | \psi ; \tau >_{\parallel}
\, d \tau ,
\label{kr1} 
\end{equation}
which is an analogon of  Eq (\ref{a_k3}) in Sec. 2. Notice that 
in contradistinction to Eq (\ref{a_k3})  mentioned, Eq (\ref{kr1}) is 
exact. (In the literature, equations of type Eq (\ref{a_k3}) are 
called "master equation" \cite{master}, or Krolikowski--Rzewuski 
equation for the distinguished component of a state vector \cite{K-R1} 
--- \cite{pra}).

Now inserting  the expected exponential form of 
$| \psi ; t>_{\parallel}$ (\ref{ex-psi1}) into Eq (\ref{kr1}) and, 
taking into account (as in Sec. 2) the fact that all 
characteristic properies of bound, or quasistationary states manifest 
themselves at times $t \gg t_{0}$, practically for $t \rightarrow 
\infty$, (here $t_{0}$ is the moment of the preparation of the 
subsystem considered), one obtains, to a very good approximation 
\begin{eqnarray}
i \frac{\partial}{\partial t} | \psi ; t>_{\parallel}  
& \cong & PHP | \psi ;t>_{\parallel} 
-  i \Big\{ \lim_{t \rightarrow \infty}
\int_{0}^{t} \Big[
PHQ
e^{\textstyle -i (t - \tau ) QHQ } QHP \times \nonumber \\
& & \hspace{1.0in} \, \times \,
e^{\textstyle  i(t - \tau )(PHP + V_{\parallel} ) } \Big]
\, d \tau  \Big\} \, 
| \psi ; t >_{\parallel} , 
\label{kr2} 
\end{eqnarray}
(where $t \gg t_{0} =0$), which is analogous to Eqs (\ref{a_k4}), Eq 
(\ref{a_k5}).

On the other hand, if the solution (\ref{psi-perp1}) of Eq 
(\ref{eq-psi-perp1}) is directly substituted into Eq (\ref{V-par}), 
then one immediatelly finds that
\begin{equation}
V_{\parallel} | \psi ;t> \stackrel{\rm def}{=} 
-i \int_{0}^{t} PHQ
e^{\textstyle -i (t - \tau ) QHQ } QHP | \psi ; \tau >_{\parallel}
\, d \tau .
\label{V=def} 
\end{equation}
Now keeping 
in mind the motivation in relation to time $t$ presented before Eq 
(\ref{kr2}) and inserting the form (\ref{ex-psi1}) 
of $|\psi ;t>_{\parallel}$ predicted by the LOY approach 
into (\ref{V=def}) one obtains the following relation which is
valid to a very good approximation for $t \gg t_{0} =0$,
\begin{eqnarray}
V_{\parallel} | \psi ;t>_{\parallel} 
& \cong & -i \Big\{  \lim_{t \rightarrow \infty }
\int_{0}^{t} \Big[
PHQ
e^{\textstyle -i (t - \tau ) QHQ } QHP \times \nonumber \\           
& & \hspace{.5in} \times \,
e^{\textstyle  i(t - \tau )(PHP + V_{\parallel} ) } 
\, d \tau  \Big\} \; 
| \psi ; t >_{\parallel}.  \label{V-par1}
\end{eqnarray}
From this equation, or from Eq (\ref{kr2}) one can infer that the 
operator (the matrix) $V_{\parallel}$ can be obtained by solving the 
nonlinear equation
\begin{equation}
V_{\parallel}  
= -i \lim_{t \rightarrow \infty }
\int_{0}^{t} PHQ
e^{\textstyle -i (t - \tau ) QHQ } QHP 
e^{\textstyle  i(t - \tau )(PHP + V_{\parallel} ) } 
\, d \tau  .   \label{V-par2}
\end{equation}
So, the consistently applied LOY method leads to the nonlinear 
equation for the effective Hamiltonian $H_{\parallel}$, 
(\ref{H-par-def1}), governing the time evolution in the subspace 
${\cal H}_{\parallel}$. Similar 
equations one can meet in theories of equations of the "master 
equation" type, \cite{master}--- \cite{pra}. 

Solutions of Eq (\ref{V-par2}) can be found, e.g., by means of the 
iteration method. Putting in (\ref{V-par2}) (see \cite{K-R1})        
\begin{equation}
V_{\parallel}^{(n + 1)}  
=  -i  \lim_{t \rightarrow \infty }
\int_{0}^{t} PHQ
e^{\textstyle -i (t - \tau ) QHQ } QHP 
e^{\textstyle  i(t - \tau )(PHP + V_{\parallel}^{(n)} ) } 
\, d \tau  ,  \label{V-par3}
\end{equation}
one can express $V_{\parallel}$ as follows
\begin{equation}
V_{\parallel} = \lim_{n \rightarrow \infty} V_{\parallel}^{(n)}.
\label{V-par4}
\end{equation}
Taking into account the fact that the contribution of the component 
$|\psi ;t>_{\perp}$ into Eq (\ref{eq-psi-par1}) for 
$| \psi ;t>_{\parallel}$ 
is (by the assumption  (\ref{weak1})) very small, and therefore that 
the matrix elements of the operator $V_{\parallel}$, (\ref{V-par}),  
expressing this contribution should be very small also, it seems 
reasonable to assume that 
\begin{equation}
V_{\parallel}^{(0)} = 0. \label{V-par(0)}
\end{equation}
Such an assumption corresponds with the similar one exploited in the 
LOY approach, i.e., which is made in Sec. 2 for the 
parameter  $\Lambda $ appearing in formulae (\ref{a_k2a}) --- 
(\ref{LOY-as}), where the final formulae for the matrix elements of 
$H_{LOY}$ were obtained by assuming that $\Lambda = 0$ (see Eq 
(\ref{a_k5})). Therefore the identification of the 
approximate solutions $V_{\parallel}^{(1)}$ of Eq (\ref{V-par3}), 
\begin{equation}
V_{\parallel}^{(1)} =
- i \lim_{t \rightarrow \infty}
\int_{0}^{t} PHQ 
e^{\textstyle -i(t - \tau )QHQ}
QHP
e^{\textstyle -i(t - \tau )PHP} \, d \tau,
\label{V(1)}
\end{equation}
with the LOY effective Hamiltonian $H_{\parallel} = H_{LOY}$, 
(or with the improved LOY effective Hamiltonian $H_{\parallel} = 
H_{LOY}^{Imp}$ ), by the relation (\ref{H-par-def1}),
\begin{equation}
H_{LOY} \; (H_{LOY}^{Imp}) \; = \; H_{\parallel}^{(1)} 
\equiv  PHP + V_{\parallel}^{(1)},  
\label{H-approx}
\end{equation}
seems to be well--grounded. 

Now let us analyse more carefuly relations (\ref{V-par}) and 
(\ref{V=def}). From these relations it follows that in fact 
the supposition made in \cite{LOY2} that 
$i \frac{\partial }{\partial t} |\psi ;t>_{\parallel}$ can be 
expressed in terms of $|\psi ; t>_{\parallel}$ only by
means of time independent coefficients, that is  that 
$V_{\parallel}$ should be time independent, 
in general is not true. Analysing the mentioned formulae
and the initial condition (\ref{psi-perp(0)})
one finds that
\begin{equation}
V_{\parallel} \equiv 
V_{\parallel}(t), \; \; \;{\rm and} \; \; \;  
V_{\parallel}(t = 0) = 0, \l; \; V_{\parallel}( t > 0) \neq 0.
\label{V(0)=0}
\end{equation}
This means that the expected in the LOY approach 
exponential form (\ref{ex-psi1}) of $|\psi ; t>_{\parallel}$
cannot be longer considered as the form which reflects accurately 
the real properties of the system considered.

Eq (\ref{eq-psi-par3}) can be solved for $V_{\parallel} = 
V_{\parallel}(t)$ to obtain
\begin{eqnarray}
|\psi ; t>_{\parallel} & = & 
e^{\textstyle -itPHP} |\psi >_{\parallel} \nonumber \\
& & +
\sum_{n=1}^{\infty} (-i)^{n}
e^{\textstyle -itPHP} 
\int_{0}^{t} dt_{1} 
\int_{0}^{t_{1}} dt_{2}
\ldots   \nonumber \\
& &
\ldots
\int_{0}^{t_{n-1}} dt_{n}
\widetilde{V_{\parallel}}(t_{1}) \cdot \ldots 
\cdot \widetilde{V_{\parallel}} (t_{n}) |\psi >_{\parallel},
\label{psi-par-real}
\end{eqnarray}
where
\begin{equation}
\widetilde{V_{\parallel}}(t)
= e^{\textstyle itPHP} V_{\parallel}(t)
e^{\textstyle -itPHP}.
\label{V-tilde}
\end{equation}

From (\ref{psi-par-real}) and (\ref{V=def}) using (\ref{V-par(0)})
one can conclude that to the lowest nontrivial order
\begin{equation}
V_{\parallel}^{(1)} (t) = -i 
\int_{0}^{t} PHQ 
e^{\textstyle -i(t - \tau )QHQ}
QHP
e^{\textstyle -i(t - \tau )PHP} \, d \tau.
\label{V(1)(t)}
\end{equation}
Note that this expression for $V_{\parallel}^{(1)}(t)$ has been
obtained from the exact de\-fi\-nition (\ref{V=def}) of 
$V_{\parallel}$ without using any assumptions considered in 
\cite{LOY1,LOY2} and leading to the formula
(\ref{V(1)}) for $V_{\parallel}^{(1)}$.

From the last formula for $V_{\parallel}^{(1)}(t)$ and from
the expression (\ref{V(1)}) for $V_{\parallel}^{(1)}$, which 
has been obtained with the use of the assumptions exploited in 
LOY papers, it follows that the conclusion following from the 
supposition made in \cite{LOY2} that $V_{\parallel}$ should be time
independent can be considered to be justified for 
$t \gg t_{0} =0$, (that is technically for $t \rightarrow \infty$).

\subsection{Assumptions leading to the standard form of $H_{LOY}$.}

Analysing the LOY derivation of the effective Hamiltonian discussed 
one can observe that the components containig the matrix elements 
$H_{kl}^{(1)}$, ($k,l =1,2$), are neglected in the right sides of 
the LOY equations equivalent to Eqs(\ref{a_k2}) in \linebreak Sec. 2 
(see Eqs (18), (19) in \cite{LOY1}, or, Eqs (A1.4), (A1.5) in 
\cite{Gaillard}, Chap. 5, Appendix 1). The analogous form of Eq 
(\ref{eq-psi-par2}) can be justified if for every finite $t \geq 0$
\begin{equation}
\parallel PH^{(1)}P | \psi ;t>_{\parallel} \parallel \; \;
\ll \;  \; \parallel PH^{(1)}Q | \psi ;t>_{\perp} \parallel .
\label{LOY-valid}
\end{equation}
(This condition replaces the earlier one (\ref{ww2}) used in Sec. 2). 
Assuming that inequality (\ref{LOY-valid})  holds, instead of Eq 
(\ref{eq-psi-par2}), to a sufficiently good approximation, one can 
consider the following equation 
\begin{equation}
i \frac{\partial}{\partial t} | \psi ; t>_{\parallel} 
\cong   m_{0}P  | \psi ;t>_{\parallel} + PH^{(1)}Q |\psi ;t>_{\perp},
\; \; (t \geq 0).
\label{eq-psi-LOY1}
\end{equation}
Next, according to the ideas leading to Eqs (\ref{eq-psi-par3}),  
(\ref{eq-psi-par4}), using (\ref{V-par}) this equation should be 
replaced by
\begin{equation}
i \frac{\partial}{\partial t} | \psi ; t>_{\parallel} 
\cong  \Big\{ m_{0}P   + V_{\parallel} \Big\}| \psi ;t>_{\parallel},
\; \; (t \geq 0). 
\label{eq-psi-LOY2}
\end{equation}
From this, one can conclude that if condition (\ref{LOY-valid}) is 
fulfilled and if the supposition adopted from \cite{LOY2} that
$V_{\parallel}$ should be time independent holds
then the solution of Eq (\ref{eq-psi-par2}) should have 
exactly the same exponential form (\ref{LOY-as}) as the solution of 
the LOY equations (18), (19) in \cite{LOY1}, (see \cite{LOY1}), 
formula (21)),
\begin{equation}
|\psi ;t>_{\parallel} \cong e^{\textstyle -it(m_{0} P + V_{\parallel})}
| \psi ;t = 0>_{\parallel} 
\equiv e^{\textstyle -it(m_{0}  + V_{\parallel}) }
| \psi >_{\parallel}. \label{psi-LOY3}
\end{equation} 
Similarly to the Eq (\ref{V-par2}) and according to the taken 
assumptions, such a form of solution of Eq 
(\ref{eq-psi-par2}) generates the suitable operator $V_{\parallel}$, 
\begin{equation}
V_{\parallel}  
= -i \lim_{t \rightarrow \infty }
\int_{0}^{t} PH^{(1)}Q
e^{\textstyle -i (t - \tau ) QHQ } QH^{(1)}P 
e^{\textstyle  i(t - \tau )(m_{0} + V_{\parallel} ) } 
\, d \tau  .   \label{V-par3LOY}
\end{equation}
This expression,  by  relations (\ref{V-par3}), (\ref{V-par(0)}),
to the lowest nontrivial order, gives
\begin{equation}
V_{\parallel}^{(1)} =  - {\Sigma} ( m_{0} ), 
\label{V-LOY}
\end{equation}
where
\begin{eqnarray}
{\Sigma} (x )  & = & PHQ \frac{1}{ QHQ - x - i0 } QHP \label{Sigma} \\
& \equiv &  PH^{(1)}Q \frac{1}{QHQ - x -i0} QH^{(1)}P . \nonumber 
\end{eqnarray}
Relations (\ref{V-LOY}) and (\ref{H-approx}) define the effective 
Hamiltonian $H_{\parallel}^{(1)}$, which coincides with $H_{LOY}$. So, 
one can write
\begin{equation}
H_{LOY} = m_{0}P - {\Sigma} ( m_{0} ).
\label{H-LOY1}
\end{equation}
From the course of the derivation of this effective Hamiltonian it 
follows that such an identification of $H_{\parallel}^{(1)}$ with 
$H_{LOY}$ is justifiable. One should stress  that such an 
approximation for $H_{LOY}$ can be considerd as sufficiently good and 
correct provided that for every $t \geq 0$ the requirement 
(\ref{LOY-valid}) holds.

There is an insignificant difference between this $H_{LOY}$ and 
$H_{LOY}^{(0)}$ (\ref{H-LOY}) derived in Sec. 2. It occurs because the 
exact solution (\ref{psi-perp1}) of Eqs 
(\ref{eq-psi-perp1}), (\ref{eq-psi-perp2}) was used, when the formula 
was derived for $V_{\parallel}$ in this Section, contrary 
to the case of $H_{LOY}^{(0)}$, where the approximate solutions 
(\ref{F_j4}) of Eq(\ref{F_j1}), corresponding to Eq 
(\ref{eq-psi-perp1}) were used.

Note that in the case of $H_{LOY}$ considered, and CPT symmetry      
conserved, assumptions (\ref{cpt1}) --- (\ref{cpt-H}) imply 
\begin{equation}
h_{11}^{LOY} = h_{22}^{LOY}, \label{h11=h22-a}
\end{equation}
where $h_{jk}^{LOY} = <{\bf j}|H_{LOY}|{\bf k}>, \; \; j,k =1,2$,i.e., 
exactly as for $H_{LOY}^{(0)}$ discussed in Sec. 2 (see 
(\ref{h11=h22})).

\subsection{Improved $H_{LOY}$.}

Let us consider in detail some implications of the main assumption of 
the LOY theory, i.e., the relation (\ref{psi-LOY3}), which is 
equivalent to (\ref{LOY-as}),  (\ref{main-as}) in Sec, 2 and (21) in 
\cite{LOY1}. This relation and similar ones are a direct consequence 
of the assumption (\ref{LOY-valid}) and the other ones of this type. 

Note that the assumption on time independence of $V_{\parallel}$ 
adopted from \cite{LOY2}, the 
relation (\ref{V-par}) and the initial condition       
(\ref{psi-perp(0)})  states that
\begin{eqnarray}
V_{\parallel} | \psi ; t = 0>_{\parallel} & = & PHQ 
| \psi ; t = 0>_{\perp}
\equiv 0,  \label{V-par=0} \\
V_{\parallel} | \psi ; t > 0>_{\parallel} & = & PHQ 
| \psi ; t > 0>_{\perp}
\neq 0.  \label{V-par=no0} 
\end{eqnarray}
The result (\ref{V-par=0}) means that neglecting the component 
$PH^{(1)}P | \psi ;t>_{\parallel}$ in the right side of Eq 
(\ref{eq-psi-LOY1}) and keeping all the remaining 
assumptions  used in \cite{LOY1,LOY2}  
lead, by the relation (\ref{psi-LOY3}), to the 
trivial form for the $| \psi ;t >_{\parallel}$:                      
\begin{equation}
| \psi ;t>_{\parallel} = e^{\textstyle -it(m_{0} P + V_{\parallel}) }
| \psi ;t = 0>_{\parallel} 
\equiv e^{\textstyle -itm_{0}  }
| \psi >_{\parallel}, \label{trivial}
\end{equation} 
which does not reflect the real processes occuring, e.g., in the 
neutral kaon complex. In other words, the assumptions of type 
(\ref{LOY-valid}), the only ones under which $H_{LOY}$ can be derived, 
force two state unstable system considered to behave as one state (one 
level) stationary subsystem. Thus,  the substitutions of type 
(\ref{psi-LOY3}) into the Eq (\ref{kr2}), or Eq (\ref{V-par1}), i.e., 
the Eq (\ref{V-par3LOY}) can not result in the approximate effective 
Hamiltonian (\ref{H-approx}) which could describe correctly the real 
properties of a two state unstable subsystem. What is more,  
if one takes into account the analysis performed at the end of
Sec. 3.1), formulae (\ref{V(0)=0}), (\ref{V(1)(t)}) then
such a conclusion seems to be quite obvious. 

A detailed analysis of the assumption (\ref{LOY-valid}) permitting
the approximate effective Hamiltonian governing the time evolution in 
two dimensional subspace of states to be of the LOY form 
(\ref{H-LOY1}) indicates that such an assumption cannot be fulfilled 
for every $t \geq 0$. One finds that at $t = 0$, and thus at $0 < t 
\rightarrow 0$ it is not satisfied. Namely, it is not consistent with 
the initial condition (\ref{psi-perp(0)}). From (\ref{psi-perp(0)}) it 
follows that $PHQ | \psi ; t=0>_{\perp} = 0$, and thus  \linebreak
$PHQ | \psi ; t \rightarrow 0>_{\perp} \simeq 0$, 
which,  if
(\ref{LOY-valid}) holds, leads to the irrational conclusion that  
at $t=0$ there should be
\[
|| PH^{(1)}P |\psi ; t = 0>_{\parallel}|| \ll 0.
\]
So, keeping in mind that 
$\parallel | \psi ;t = 0>_{\parallel} \parallel = 1$ one concludes 
that  there must be
$\parallel PH^{(1)}P | \psi ;t \rightarrow 0>_{\parallel} \parallel 
\; >  \; \parallel PH^{(1)}Q | \psi ;t \rightarrow  0>_{\perp}  \parallel 
\simeq 0$ for $0 <t \rightarrow 0$ instead of (\ref{LOY-valid}).  
If (\ref{psi-perp1}) is inserted into (\ref{LOY-valid}) then one can
see that the condition (\ref{LOY-valid}) cannot be fulfilled for
$ t \gg t_{0} = 0$ either.   This 
means that the derivation of $H_{LOY}$ is incoherent. (The same 
conclusion refers to all derivated formulae for the LOY effective 
Hamiltonian in the literature, including \cite{LOY1} --\cite{4}, 
where the approximations equivalent to the assumption 
(\ref{LOY-valid}) were used). 
On the one hand, in the LOY treatment of time evolution in a two state 
subspace initial conditions are defined for $t = t_{0} \equiv 0$ and 
solutions of approximate equations of Eq (\ref{eq-psi-LOY1}) type are 
discussed for $t \geq t_{0} = 0$, up to $t = + \infty$. On the other 
hand, within this treatment the approximation of type (\ref{LOY-valid})  
is used and this approximation in not true for the whole domain of the 
parameter $t$, but only for its part (for $t \gg t = t_{0} =0$). In 
other words, conditions of the (\ref{LOY-valid}) type can never reflect 
the real properties of time evolution in the two state subsystem 
considered. Therefore $H_{LOY}$ derived within the use of this condition 
and the assumption of time independece of $V_{\parallel}$ 
is unable to describe correctly all 
the real properties of the system under considerations. 

The defects of the LOY method described above can be easily rectified. 
It is sufficient to abandon this questionable condition 
(\ref{LOY-valid}). In other words, instead of approximate equations of  
type (\ref{eq-psi-LOY1}) one should use equations of the type 
(\ref{eq-psi-par1}), (\ref{eq-psi-par2}) containing component 
$PH^{(1)}P | \psi ,t>_{\parallel}$, (or, matrix elements $H_{jk}$, 
$(j,k =1,2)$  in the case of equations of the type (\ref{a_k1})).  
Thus, the exponential form of $| \psi ,t>_{\parallel}$ given by the 
relation (\ref{psi-LOY3}) cannot be considered at all, 
but only the the formula (\ref{V(1)(t)}) for $V_{\parallel}(t)$ 
should be used. 

So, let us use  the above mentioned improvements of the LOY method and 
find the approximate $V_{\parallel}^{(1)} \equiv V_{\parallel}^{Imp}$ 
by means of the formula (\ref{V(1)}) for $PHP = m_{0}P + PH^{(1)}P$, 
which can be derived from (\ref{V(1)(t)}).
In such a case one finds
\begin{eqnarray}
V_{\parallel}^{(1)}   \equiv V_{\parallel}^{Imp} 
& = & -i  \lim_{t \rightarrow \infty }
\int_{0}^{t} \Big\{
PH^{(1)}Q e^{\textstyle -i (t - \tau ) (QHQ  - m_{0}) } QH^{(1)}P 
\times \nonumber \\& & \hspace{1.0in} \times 
e^{\textstyle  i(t - \tau )PH^{(1)}P } 
\,  \Big\} d \tau  .  \label{V-imp1}
\end{eqnarray}

To evaluate this integral it is necessary to calculate $\exp [it 
PH^{(1)}P]$. Keeping in mind that $PH^{(1)}P$ is the hermitian $(2 
\times 2)$ matrix and using the Pauli matrices representation
\begin{equation}
PH^{(1)}P \equiv h_{0}^{(1)} I_{\parallel} +  {\bf  h}^{(1)}
\cdot {\bf s},
\label{H1-pauli}
\end{equation}
where ${\bf h}^{(1)}$ and ${\bf s}$ denote the following vectors:
${\bf h}^{(1)} = (h_{x}^{(1)}, h_{y}^{(1)},  h_{z}^{(1)})$,
${\bf s}$ = (${\sigma}_{x}, {\sigma}_{y}, {\sigma}_{z}$), and
$I_{\parallel}$ is the unit operator in ${\cal H}_{\parallel}$, and,
of course, $I_{\parallel} \equiv P$,
\begin{equation}
{\bf h}^{(1)} \cdot {\bf s}  =  h_{x}^{(1)} {\sigma}_{x} +
h_{y}^{(1)} {\sigma}_{y} + h_{z}^{(1)} {\sigma}_{z} ,
\label{H1-pauli1}
\end{equation}
\[
h_{0}^{(1)}  =  \frac{1}{2} [ H_{11}^{(1)} + H_{22}^{(1)} ] ,
\]
\[
h_{z}^{(1)}  =  \frac{1}{2} [ H_{11}^{(1)} - H_{22}^{(1)} ] ,
\]
\begin{eqnarray*}
({\kappa}^{(1)} )^{2}  \stackrel{\rm def}{=}  {\bf h}^{(1)} \cdot {\bf 
h}^{(1)}& = &  (h_{x}^{(1)} )^{2} + (h_{y}^{(1)} )^{2} + 
(h_{z}^{(1)} )^{2}   \\
&\equiv & H_{12}^{(1)} H_{21}^{(1)} + (h_{z}^{(1)} )^{2} ,
\end{eqnarray*}
(${\sigma}_{k}$,  ($k = x,y,z$), are the Pauli matrices), 
one finds
\begin{equation}
e^{\textstyle  \pm  itPH^{(1)}P } 
= e^{\textstyle  \pm ith_{0}^{(1)} }
\Big[ I_{\parallel} \cos (t{\kappa}^{(1)}) \pm 
i \frac{ {\bf h}^{(1)} \cdot {\bf s} }{{\kappa}^{(1)} } \sin 
(t{\kappa}^{(1)}) \Big] .\label{ex-H1}
\end{equation}
It is conveniet to use (\ref{H1-pauli}) again and replace 
${\bf h}^{(1)} 
\cdot{\bf s}$ by ${\bf h}^{(1)} \cdot{\bf s} = PH^{(1)} P - 
h_{0}^{(1)} P$ in Eq (\ref{ex-H1}), which, after some algebra, gives
\begin{eqnarray}
e^{\textstyle  + itPH^{(1)}P } & \equiv &
\frac{1}{2} e^{\textstyle it( h_{0}^{(1)} +{\kappa}^{(1)} )}
[(1 - \frac{h_{0}^{(1)} }{{\kappa}^{(1)} }) P + 
\frac{1}{{\kappa}^{(1)} } 
PH^{(1)}P ] \nonumber \\
& + & \frac{1}{2} e^{\textstyle it( h_{0}^{(1)} -{\kappa}^{(1)} )}
[(1 + \frac{h_{0}^{(1)} }{{\kappa}^{(1)} }) P - 
\frac{1}{ {\kappa}^{(1)} } 
PH^{(1)}P ] .  \label{ex-H1b}
\end{eqnarray}
Now, inserting (\ref{ex-H1b}) into (\ref{V-imp1}) yields             
\begin{eqnarray}
V_{\parallel}^{Imp} & = & - \frac{1}{2} \Sigma (m_{0} + h_{0}^{(1)} + 
{\kappa}^{(1)} ) \Big[ (1 - \frac{h_{0}^{(1)}}{{\kappa}^{(1)}} )P +
\frac{1}{{\kappa}^{(1)}} PH^{(1)} P \Big] \nonumber \\
& \; & - \frac{1}{2} \Sigma (m_{0} + h_{0}^{(1)} - 
{\kappa}^{(1)} ) \Big[ (1 + \frac{h_{0}^{(1)}}{{\kappa}^{(1)}} )P -
\frac{1}{{\kappa}^{(1)}} PH^{(1)} P \Big] . \label{V-imp2} 
\end{eqnarray}
This means (by (\ref{H-approx}) ) that the improved LOY method leads 
to the following effective Hamiltonian $H_{LOY}^{Imp}$ governing the 
time evolution in the two state subspace,
\begin{equation}
H_{LOY}^{Imp} = m_{0} P + PH^{(1)} P + V_{\parallel}^{Imp}.
\label{H-LOY-imp}
\end{equation}
This effective Hamiltonian $H_{LOY}^{Imp}$ differs significantly 
from the standard expression (\ref{H-LOY}) for $H_{LOY}^{(0)}$ and 
from (\ref{H-LOY1}).  The properties of the matrix elements of these 
effective Hamiltonians, both of which are calculated for the CPT 
invariant system (\ref{cpt-H}), (\ref{cpt-H1}), are the main and the  
most conspicuous  difference. This main difference can be found by 
comparing standard formula (\ref{h-LOY-jk}) for matrix elements 
$h_{jk}^{LOY(0)}$ of $H_{LOY}^{(0)}$ with the formulae for matrix 
elements $h_{jk}^{Imp}$ of $H_{LOY}^{Imp}$,
\begin{equation}
h_{jk}^{Imp} = <{\bf j}|H_{LOY}^{Imp}|{\bf k}> =
m_{0} {\delta}_{jk} + H^{(1)}_{jk} + v_{jk}^{Imp}, \; \; 
({\scriptstyle j,k =1,2}),\label{h-jk-imp}
\end{equation}
where, 
\begin{eqnarray}
v_{j1}^{Imp} = & - & \frac{1}{2} \Big( 1 + 
\frac{h_{z}^{(1)}}{{\kappa}^{(1)}} \Big){\Sigma}_{j1} (m_{0} + 
h_{0}^{(1)} +  {\kappa}^{(1)} ) 
\label{v-j1-imp1} \\
& - & \frac{1}{2} \Big( 1 - 
\frac{h_{z}^{(11)}}{{\kappa}^{(1)}} \Big){\Sigma}_{j1} (m_{0} + 
h_{0}^{(1)} - {\kappa}^{(1)} )  \nonumber  \\
& - & \frac{H_{21}^{(1)}}{2 
{\kappa}^{(1)}} {\Sigma}_{j2} (m_{0} + h_{0}^{(1)} + {\kappa}^{(1)} )+ 
\frac{H_{21}^{(1)}}{2 {\kappa}^{(1)}} {\Sigma}_{j2} (m_{0} + 
h_{0}^{(1)} - {\kappa}^{(1)} ) , \nonumber 
\end{eqnarray}
\begin{eqnarray}
v_{j2}^{Imp} = & - & \frac{1}{2} \Big( 1 - 
\frac{h_{z}^{(1)}}{{\kappa}^{(1)}} \Big){\Sigma}_{j2} (m_{0} + 
h_{0}^{(1)} + {\kappa}^{(1)} ) 
\label{v-j2-imp1} \\
& - & \frac{1}{2} \Big( 1 + 
\frac{h_{z}^{(1)}}{{\kappa}^{(1)}} \Big){\Sigma}_{j2} (m_{0} + 
h_{0}^{(1)} - {\kappa}^{(1)} )  \nonumber  \\
& - & \frac{H_{12}^{(1)}}{2 
{\kappa}^{(1)}} {\Sigma}_{j1} (m_{0} + h_{0}^{(1)} + {\kappa}^{(1)} )
+ \frac{H_{12}^{(1)}}{2 {\kappa}^{(1)}} {\Sigma}_{j1} 
(m_{0} + h_{0}^{(1)} - {\kappa}^{(1)} ) , \nonumber
\end{eqnarray}
${\Sigma}_{jk} ( \varepsilon )
= <  {\bf j} \mid \Sigma ( \varepsilon ) \mid{\it {\bf k}} >$, 
and $j,k =1,2$. Now, using (\ref{cpt-H0}) --- (\ref{cpt-H}) it is 
not difficult to conclude from (\ref{h-jk-imp}) --- (\ref{v-j2-imp1}) 
that for the CPT invariant but CP noninvariant system, it must be
\begin{equation}
[ \Theta , H] = 0 \; \; \Rightarrow \; \; 
h_{11}^{Imp} \neq h_{22}^{Imp}, \label{h11-h22-no}
\end{equation}
contrary to the standard LOY result (\ref{h11=h22}). It should be 
emphasized in this place  that improving the LOY method, only 
the consistency of the initial conditions (\ref{init}) and 
({\ref{psi-perp(0)}) (or (\ref{a(0)}) and (\ref{F(0)}) )  for the 
problem with the approximations used (\ref{weak1}), (\ref{ex-psi2}) 
and with the geometry (the dimension) of ${\cal H}_{\parallel}$
has been taken into account much more rigorously than it was made 
by Lee, Oehme and Yang. 
All steps leading to the 
formulae for $H_{LOY}^{Imp}$  are well founded and do not impair the 
main ideas of the standard LOY method. So, the $H_{LOY}^{Imp}$ should 
reflect the real properties of the system considered much better than 
it is possible within the use of the standard LOY effective 
Hamiltonian (\ref{h-LOY-jk}), (\ref{H-LOY}).  In this context, the 
result (\ref{h11-h22-no}) seems to have serious consequences when 
interpreting CPT invariance tests, e.g., for the neutral kaon complex 
\cite{cpt-qm}. 

\section[4]{Effective Hamiltonian $H_{\parallel}$ for three state \\ 
complex.}

Using the LOY method the  effective Hamiltonian $H_{\parallel}$ 
governing the time evolution in $n$--dimensional subspace 
${\cal H}_{\parallel}$ of 
state space $\cal H$ for $n > 2$ can also be found. A derivation of 
such a $H_{\parallel}$ is rather time consuming  when one uses the 
standard LOY approximation and considers equations of type 
(\ref{a_k1}), (\ref{F_j1}) for amplitudes $a_{j}(t),\; (j = 1,2, 
\ldots , n)$, (\ref{psi-gen}). On the 
other hand, such a purpose can be realized relatively easy if one  
applies the improved LOY method used in Sec. 3 and uses Eqs 
(\ref{eq-psi-par1}), (\ref{eq-psi-perp1}) for components 
$| \psi ;t>_{\parallel} $, 
$| \psi ;t>_{\perp}$, (\ref{psi-P}), (\ref{psi-Q}), of a state vector 
$|\psi ;t> \in {\cal H}$ instead of the mentioned equations for 
amplitudes $a_{j}(t)$. These equations together with the initial 
condition (\ref{psi-perp(0)}) and  assumptions (\ref{weak1}), 
(\ref{V-par}) lead to the Equation (\ref{V-par2}) for $V_{\parallel}$ 
and thus, by (\ref{V-par3}), (\ref{V-par(0)}), (similarly to the case 
considered in Subsection 3.3),   to the approximate formula 
(\ref{H-approx}) for the improved effective Hamiltonian 
$H_{\parallel}^{(1)}$. In the case of a three level subsystem this 
effective Hamiltonian will be denoted as $H_{\parallel}^{(1)} \equiv 
H_{\parallel}^{(3d)}$, and for the operator $V_{\parallel}^{(1)}$ 
defining the $H_{\parallel}^{(3d)}$ the symbol $V_{\parallel}^{(3d)}$ 
will be used. Considering the general case described by Eqs 
(\ref{eq-psi-par1}), (\ref{eq-psi-perp1}), and using (\ref{V(1)}) 
one finds 
\begin{eqnarray}
V_{\parallel}^{(1)}   \equiv V_{\parallel}^{(3d)} 
& = & -i  \lim_{t \rightarrow \infty }
\int_{0}^{t} \Big\{
PHQ e^{\textstyle -i (t - \tau ) QHQ } QHP \times \nonumber \\
&  & \hspace{1.0in}  \times \,
e^{\textstyle  i(t - \tau )PHP } 
\,  \Big\} d \tau  ,  \label{V-3d1}
\end{eqnarray} 
and thus (according to (\ref{H-approx}) )
\begin{equation}
H_{\parallel}^{(3d)} = PHP + V_{\parallel}^{(3d)}. 
\label{H-3d1}
\end{equation}
So,  the only problem is to calculate $\exp [itPHP]$ in (\ref{V-3d1}) 
for the case of $\dim ({\cal H}_{\parallel}) = 3$. 

Let the subspace ${\cal H}_{\parallel}$ be spanned by a set of 
orthonormal vectors \linebreak $\{ |{\bf e}_{j}>{\}}_{j=1,2,3.} 
\in {\cal H}$, $<{\bf e}_{j}|{\bf e}_{k}> ={\delta}_{jk}$. Then the 
projection operator P defining this subspace (see (\ref{Hu}) ) can be 
expressed as follows
\begin{equation}
P = \sum_{j=1,2,3} |{\bf e}_{j}><{\bf e}_{j}| \equiv 
I_{\parallel}^{(3d)},
\label{P-3d}
\end{equation}
where $I_{\parallel}^{(3d)}$ is the unity for the three dimensional 
subspace ${\cal H}_{\parallel}$ considered, 
and the complementary projector $Q$, (\ref{Q}), equals $Q = I - P$. 
 
The operator $PHP$ is selfadjoint, so the $(3 \times 3)$ matrix 
representing $PHP$ in the subspace ${\cal H}_{\parallel}$ is Hermitian 
matrix. Solving  the eigenvalue problem for this matrix,
\begin{equation}
PHP |{\lambda}_{j}> = {\lambda}_{j} |{\lambda}_{j}>, \; \;
({\scriptstyle j =1,2,3}),
\label{l-j}
\end{equation}
one obtains the eigenvalues ${\lambda}_{j} =  {\lambda}_{j}^{\ast}$, 
and eigenvectors $|{\lambda}_{j} >$, $(j =1,2,3)$. For 
simplicity we assume that ${\lambda}_{1} \neq {\lambda}_{2}  \neq
{\lambda}_{3} \neq {\lambda}_{1}$, i.e., that all $ |{\lambda}_{j} >$ 
are orthogonal,
\begin{equation}
<{\lambda}_{j} |{\lambda}_{k} > 
= <{\lambda}_{j} |{\lambda}_{j} > {\delta}_{jk}, \; \; 
({\scriptstyle j,k =1,2,3}).
\label{l-jk}
\end{equation}
By means of these eigenvectors one can define new projection operators,
\begin{equation}
P_{j} \stackrel{\rm def}{=} 
\frac{1}{<{\lambda}_{j} |{\lambda}_{j} >}
|{\lambda}_{j}><{\lambda}_{j}|, \; \;
({\scriptstyle j =1,2,3}).
\label{P-j}
\end{equation}
The property (\ref{l-jk}) of the solution of the eigenvalue problem 
for $PHP$ considered implies that
\begin{equation}
P_{j} P_{k} = P_{j} {\delta}_{jk}, \; \;
({\scriptstyle j =1,2,3}), 
\label{P-jk}
\end{equation}
and that the completeness requirement for the subspace 
${\cal H}_{\parallel}$ 
\begin{equation}
\sum_{j=1,2,3} P_{j} = P, 
\label{copm}
\end{equation}
holds.
Now, using the projectors $P_{j}$ one can write
\begin{equation}
PHP = \sum_{j=1,2,3} {\lambda}_{j} P_{j}, 
\label{PHP-sp}
\end{equation}
and
\begin{equation}
P e^{\textstyle +itPHP} = P \sum_{j=1,2,3} 
e^{+it {\lambda}_{j} }P_{j}.
\label{ex-PHP-sp}
\end{equation}

This last relation is the solution for the problem of finding 
$\exp [itPHP]$ and leads to the following formula for                
$V_{\parallel}^{(3d)}$,
\begin{equation}
V_{\parallel}^{(3d)} 
=  -i  \lim_{t \rightarrow \infty }
\sum_{j=1,2,3} \int_{0}^{t}
PHQ e^{\textstyle -i (t - \tau ) (QHQ - {\lambda}_{j})} QHP 
\,   d \tau  \, P_{j} . \label{V-3d2}
\end{equation}
A computation  of the value of this integral can be easy performed and 
yields
\begin{equation}
V_{\parallel}^{(3d)} = - \sum_{j=1,2,3}
\Sigma ({\lambda}_{j}) P_{j},
\label{V-3d-fin}
\end{equation}
(where $\Sigma ( \lambda )$ is defined by the formula (\ref{Sigma}) ), 
which by (\ref{H-3d1}) solves the problem of finding the improved LOY 
effective Hamiltonian governing the time evolution in the three state 
subspace ${\cal H}_{\parallel}$ of the total state space $\cal H$. 

The results obtained in this Section can be easily generalized to the 
case of $\dim ({\cal H}_{\parallel}) = n > 3$. 

\section[5]{Final remarks.}

Detailed analysis of assumptions leading to the standard form of the 
LOY effective Hamiltonian governing the time evolution in a two state 
subsystem indicates that some assumptions, which have been used in 
the LOY treatment of the problem, and which the WW theory of single 
line width uses, should not be directly applied to the case of two, 
or more, level subsystems interacting with the rest of the physical 
system considered. Namely, when one considers the single line width 
problem in the WW manner it is quite sufficient to analyse the 
smallness of matrix elements of the interaction Hamiltonian, 
$H^{(1)}$, only. For the multilevel problem,
contrary to the single line problem, such 
a smallness does not ensure the suitable smallness of components of 
the evolution equations containing these matrix elements. 
Moreover, there is no necessity of taking 
into account the internal dynamics of the subsystem, which also has 
an effect on the widths of levels in many levels subsystems, in such a 
case. The observed level widths in two and more level 
subsystem depend on the interactions of this subsystem with the rest, 
but they  also depend on the interactions between the levels forming 
this subsystem. So, the internal interactions in the subsystem 
considered cannot be neglected when one wants to describe the real 
properties of multi state subsystems. 

From the form of Eqs (18) --- (20) in 
\cite{LOY1} (or, Eqs (A1.) --- (A1.6) in \cite{Gaillard}, Appendix 1 
of Chap. 5) it follows that the LOY and related treatments of time 
evolution in two state subsystem use the WW theory of the single line 
width without any modification of the questionable points of the WW 
method and do not consider at all the aspects of time evolution in 
many state subsystem mentioned above. 
When one  wants to apply the LOY method of  searching for the 
properties of the time evolution in a two level subsystem,  
in order to be more rigorous than it was done in \cite{LOY1} --- 
\cite{baldo} and than it is possible within the standard WW 
approach, one should replace requirements (\ref{ww1b}) --- 
(\ref{ww3}) by the following ones
\begin{eqnarray}
| \sum_{J, \varepsilon} F_{J}( \varepsilon ;t)
H_{kJ}^{(1)}( \varepsilon )| &\ll& m_{0} |a_{k}(t)|,
\; \; \; ({\scriptstyle k =1,2}), \label{ww1br} \\
| \sum_{l=1,2}
H_{kl}^{(1)} a_{l}(t)| &\ll& | \sum_{J, \varepsilon} F_{J}
( \varepsilon ;t)
H_{kJ}^{(1)}( \varepsilon )|, 
\; \; \; ({\scriptstyle k =1,2}), \label{ww2r}
\end{eqnarray}
and 
\begin{equation}
|\sum_{L, \varepsilon '} F_{L}( \varepsilon ';t)
H_{J,L}^{(1)}( \varepsilon ,\varepsilon ')| 
\ll | \sum_{k=1,2}
H_{Jk}^{(1)}( \varepsilon ) a_{k}(t)| .
\label{ww3r} 
\end{equation} 
Such a form of assumptions replacing (\ref{ww1b}) --- (\ref{ww3}) 
enable, e.g., to detect the inconsistencies between the main LOY 
assumption (\ref{main-as}) (or, (\ref{LOY-as}) ) and  the initial 
condition (\ref{F(0)}).  From (\ref{F(0)}) it follows that the 
requirments of type (\ref{ww2r}), the only ones 
under which the approximate Eqs (\ref{a_k2}) are 
sufficiently accurate, can not be fulfilled for $t=0$ and for $t 
\rightarrow 0$. (It is impossible to draw a similar conclusions from 
the assumptions of type (\ref{ww2}) ). So, the expected and assumed 
exponential form, (\ref{LOY-as}), of $| \psi ;t>_{\parallel}$, 
(\ref{psi-par-t}), should take into account the fact that for short 
$t$ the influence of $H^{(1)}$ on the form of $| \psi ;t>_{\parallel}$ 
predominates over the the influence coming from the component 
containing $\sum_{J, \varepsilon} F_{J}( \varepsilon ;t)$ in Eq 
(\ref{a_k1}). The influence of this last component can become crucial 
only for suficiently large times $t \geq T_{as} >0$. It seems to be 
obvious that $| \psi ;T_{as}>_{\parallel} \neq | \psi 
;t= t_{0} =0>_{\parallel}$. So, whether one should replace $| \psi 
;t=0>_{\parallel}$ by $| \psi ;t = T_{as}>_{\parallel}$ in the 
assumption (\ref{LOY-as}), or one should leave 
$| \psi ;t=0>_{\parallel}$ unchanged in (\ref{LOY-as}) but 
change the index of the power in (\ref{LOY-as}) adding  $H^{(1)}$, 
cut down to the subspace ${\cal H}_{\parallel}$, there. 
These cases, similarly to the improved LOY method used in Subsection 
3.3, lead to the effective Hamiltonian 
$H_{\parallel} = H_{LOY}^{Imp}$, which differes from the standard LOY 
effective Hamiltonian $H_{LOY}^{(0)}$, (\ref{h-LOY-jk}), (\ref{H-LOY}). 

Analysing the standard derivation of $H_{LOY}$ \cite{LOY1,Gaillard} 
one can draw a conclusion which seems to be strange 
that conditions of type (\ref{ww1b}) --- (\ref{ww3}), necessary to 
obtain this $H_{LOY}$, lead to the same form of the efective 
Hamiltonian, $H_{\parallel}$, 
governing the time evolution in subspace ${\cal H}_{\parallel}$ 
independently of the dimension of this ${\cal H}_{\parallel}$. 
This means that the properties of the subsystem considered which 
manifests themselves during the time evolution, should not depend on 
the dimension of the subspace of states of this subsystem. 
The common form of $H_{\parallel}$
is given by (\ref{H-LOY}) and (\ref{H-LOY1}) and this is the form 
which can be obtained by means of the improved LOY method only for 
one--dimensional subspace ${\cal H}_{\parallel}$. Taking
\begin{equation}
P \equiv | \psi >< \psi | \stackrel{\rm def}{=} P_{\psi},
\; \; Q = I - P_{\psi},
\label{p-1dim}
\end{equation}
where $< \psi | \psi > = 1$, one has 
\begin{equation}
PHP = < \psi |H| \psi > P_{\psi},
\label{php-1dim}
\end{equation}
and thus, using (\ref{V(1)}) one can calculate 
$V_{\parallel}^{(1)} \stackrel{\rm def}{=} V_{\parallel}^{(1d)}$, 
which equals
\begin{equation}
V_{\parallel}^{(1d)} = - {\Sigma}_{\psi} (< \psi |H| \psi >),
\label{V-par-1d}
\end{equation}
where ${\Sigma}_{\psi}(x)$ is defined by the relation (\ref{Sigma}) 
for $P \equiv P_{\psi}$. So, the approximate effective Hamiltonian, 
$H_{\parallel}^{(1)} \stackrel{\rm def}{=} H_{\parallel}^{(1d)}$, for 
the case $\dim ({\cal H}_{\parallel}) = 1$ appears to be (see 
(\ref{H-approx}) ),
\begin{equation}
H_{\parallel}^{(1d)} =  P_{\psi} H P_{\psi} +
V_{\parallel}^{(1d)} .
\label{H-par-1d}
\end{equation}
Such a form of $V_{\parallel}^{(1)}$, and thus of the effective 
Hamiltonian $H_{\parallel}^{(1)}$, is produced by the standard LOY 
approach for the case of the arbitrary dimension of 
${\cal H}_{\parallel}$. From the last 
formula and from the relations (\ref{V-3d-fin}), (\ref{H-3d1}), and 
(\ref{H-LOY-imp}) it follows that
the form of the effective Hamiltonians obtained within  the use of the 
improved LOY method desribed in Sec. 3 depends on the geometry of the 
problem, i.e., on the dimension of the subspace 
${\cal H}_{\parallel}$.  Such an implication of 
the improved LOY method, (contrary to the result, which can be 
obtained by the standard LOY method), 
seems to be quite natural and obvious for the real physical systems. 
Therefore the improved LOY method, which is consistent with the 
initial condition for the problem, (\ref{F(0)}) or 
(\ref{psi-perp(0)}), and more rigorous than the standard one, should 
reflect the real properties of the system considered more accurately 
than it is possible within the use of the LOY theory in its original 
form. 

It seems that the differences between $H_{LOY}$ and $H_{LOY}^{Imp}$ 
can be explained by means of the resolvent formalism. From the point
of view of this formalism $H_{LOY}$ is generated by the pole 
approximation (see \cite{Bil-Kab,beyond,Cohen} and the paper by 
Horwitz and Marchand cited in \cite{master}). When the 
nonorthogonality of the residues in neglected then within this
approximation one obtains the effective Hamiltonian $H_{LOY}$,
that is, in general, the effective Hamiltonian of the form
(\ref{V-par-1d}), (\ref{H-par-1d}).
In fact, as it has been pointed out in \cite{Cohen} these residues 
are not orthogonal. Taking into account this fact results in
the effective Hamiltonian for the two--channel problem which
differs from the standard form of $H_{LOY}$. The connection
between the approach based on the more exact resolvent formalism
and the method described at the beginning of Sec. 3.1) and at the 
end of that Subsection can be found by the use of the Laplace
transforms. Eq. (\ref{kr1}) and the relation (\ref{V=def}) 
are the exact ones. There is a one--to--one relation between
Eq. (\ref{kr1}) for $|\psi ; t>_{\parallel}$ and Equation 
for $| \widetilde{\psi}; z>_{\parallel} = 
{\cal L}[|\psi ;t>_{\parallel}](z)$ (here ${\cal L}[.](z)$ denotes  
the Laplace transform) in terms of the reduced resolvent. 
Solution (\ref{psi-par-real}) 
of the Eq. (\ref{eq-psi-par3}) for $|\psi ;t>_{\parallel}$ can    
also be considered as the exact one. Formula (\ref{V(1)(t)})
for $V_{\parallel}^{(1)}(t)$, from which the expression 
(\ref{V(1)}) for the  $V_{\parallel} \simeq V_{\parallel}^{(1)}$
follows, has been obtained using these exact relations. 
On the other hand, $V_{\parallel}^{(1)}$ has  been used Sec. 3.3)
to obtain $H_{LOY}^{Imp}$. Therefore the  supposition made above
that the nonorthogonality of residues is responsible 
for the difference between the forms of $H_{LOY}$ and
$H_{LOY}^{Imp}$ seems to be justified. 

Note that, as it has been shown in Sec. 4, the discussed improved 
method allows one to relatively easy compute  the effective 
Hamiltonian $H_{\parallel}$ for $n$--dimensional ($n \geq 2$) 
subspace ${\cal H}$ of states.

The size of the effect (\ref{h11-h22-no}) taking place for 
$H_{LOY}^{Imp}$ can be easily estimated for the generalized 
Fridrichs--Lee model (see Appendix). Within this model one can obtain 
that
\begin{equation}
(h_{11}^{Imp}  -  h_{22}^{Imp})
\simeq 0,94  \times 10^{-14} {\rm Im}\,(H_{12}),
\label{FL0}
\end{equation}
where $H_{jk} = <{\bf j}|H|{\bf k}>, \; (j,k = 1,2)$. Comparing
this estimation and, e.g., the limit 
$|m_{K_{0}} - m_{\overline{K}_{0}}| < 2,0 \times 10^{-18}
|m_{K_{0}}|$ (see formula (121) in \cite{fermilab}) one  
can conclude that the $H_{LOY}^{Imp}$ does not lead to effects
which are in conflict with the results of recent experiments.

From (\ref{FL0}) it follows that the effect (\ref{h11-h22-no}) is 
very small indeed, and it is beyond the
accuracy of today's experiments with neutral kaons. Test of 
higher accuracy are expected to be 
performed in the near future \cite{dafne}. So, there is a chance that 
these tests will confirm this effect.
Nevertheless, the improved formulae for the LOY effective Hamiltonian 
seem to have a great meaning for the interpretation of some recent 
theoretical speculations such as those considered, for instance, in 
\cite{cpt-qm,5th-f,sr-eqp}.  Indeed, the parameters 
used in \cite{cpt-qm} to describe the deviations of quantum 
mechanics, or violations of CPT, are of similar order to  
(\ref{FL0}). This means that the interpretation of CPT tests, or tests 
of modified quantum mechanics, based on the theory developed in 
\cite{cpt-qm} may be incorretct. A similar conclusion seems to be 
right with reference to theories describing effects of external fields 
on the neutral kaon system \cite{5th-f}.  Also, the interpretation of 
tests of special relativity and of the equivalence principle 
\cite{sr-eqp} is based on the standard form, (\ref{H-LOY}), of the 
$H_{LOY}^{(0)}$. The order of the effects discussed in \cite{sr-eqp} 
can be compared to (\ref{FL0}). So it seems to be obvious that  the 
application of $H_{LOY}^{Imp}$, (\ref{H-LOY-imp}) instead of 
$H_{LOY}^{(0)}$, when one considers theories developed in all these 
papers, can lead to conclusions which need not agree with 
those obtained in \cite{cpt-qm,5th-f,sr-eqp}. 

The last observation is that 
comparing the formulae for the matrix elements, (\ref{h-jk-imp}),
of the improved LOY effective Hamiltonian, $H_{LOY}^{Imp}$, with the 
formulae for the matrix elements of the effective Hamiltonian 
$H_{\parallel} \stackrel{\rm def}{=}H_{\parallel}(t \rightarrow 
\infty)$  derived from the Krolikowski--Rzewuski equation 
\cite{K-R1} --- \cite{K-R-ur} in \cite{ur2} and discussed also in 
\cite{ur1}, one finds that all they  are identical. Also, the general 
formula, (\ref{V-par2}), (\ref{V-par3}),  for the operator 
$V_{\parallel}^{(1)}$ is simply the asymptotic case of the formula 
for $V_{\parallel}^{(1)}(t)$ obtained in \cite{ur2,pra}, namely 
$V_{\parallel}^{(1)} \equiv \lim_{t \rightarrow 
\infty}V_{\parallel}^{(1)}(t)$. So, the formalism applied in 
\cite{ur1,ur2} and also in \cite{is} to describe the properties of 
the neutral kaon and similar complexes, should not be considered as an 
alternative approach to the description of time evolution in such 
complexes. Simply, the formalism mentioned is 
more rigorous than the improved LOY method, but both these approaches 
produce the same formulae for the approximate effective Hamiltonians 
for the problem. \\
\hfill \\ \hfill \\

\renewcommand{\theequation}%
{\Alph{section}.\arabic{equation}}
\appendix
\section{Appendix}
\setcounter{equation}{0}
In 
the generalized Fridrichs--Lee 
model, the Hamiltonian is given by (see (2.1) in \cite{chiu}, or 
(3.19) in \cite{chiu1}).  
\begin{eqnarray}
H = \sum_{j,k=1}^{2} m_{jk} V_{j}^{+}V_{k} &+&
\sum_{n=1}^{N} {\mu}_{n} N_{n}^{+}N_{n}
+ \int_{0}^{\infty} \, d \omega \, {\Theta}^{+}(\omega ) \Theta
(\omega ) \nonumber \\
&+& 
\int_{0}^{\infty} \, d \omega \, \sum_{j,n} \, g_{jn}(\omega )
V_{j} N^{+}_{n} {\Theta}^{+}(\omega )  \label{H-FL} \\
&+& 
\int_{0}^{\infty} \, d \omega \, \sum_{j,n} \, g_{nj}(\omega )
V_{j}^{+} N_{n} {\Theta}(\omega ) , \nonumber
\end{eqnarray}
where $g_{nj}(\omega ) = g^{\ast}_{jn}(\omega )$. The bare particles 
are $V_{1},V_{2}, N_{n} \; (1 \leq n \leq N)$, and $\Theta$ particles. 
The following "charges" are conserved in this model:
\begin{eqnarray*}
Q_{1} &=& \sum_{j=1}^{2} V^{+}_{j}V_{j} + 
\sum_{n=1}^{N} N^{+}_{n}N_{n} , \\
Q_{2}  &=&
\sum_{n=1}^{N} N^{+}_{n}N_{n} -
\int \, d \omega {\Theta}^{+}(\omega ) \Theta (\omega ) .
\end{eqnarray*}
The corresponding eigenvalues will be denoted by $q_{1}$ and $q_{2}$. 
The Hilbert space in this model is devided into orthogonal sectors 
${\cal H}(q_{1},q_{2})$, each with different assignment of 
$q_{1}$ and $q_{2}$  values. Considering the lowest non\-tri\-vial 
sector, where $q_{1} =1$ and $q_{2} = 0$ and the bare states are 
labeled by $|V_{j}> = V_{j}^{+} |0>, \, (j =1,2), \, |n, \omega > =
N^{+}_{n} {\Theta}^{+} (\omega ) |0>, \, (n = 1,2, \ldots , N$),
and then identifying $V_{1}$ as $K_{0}$ and $V_{2}$ as 
${\overline{K}}_{0}$, after some algebra,  one finds \cite{ur2}
\begin{eqnarray}
h_{11}^{Imp}  -  h_{22}^{Imp}  & =  &
\frac{i}{4}
\frac{ m_{21} {\Gamma}_{12} - 
m_{12}{\Gamma}_{21} }{|m_{12}|} \times  \label{FL1} \\
& & \times 
\Big\{
\frac{(m_{0} - \mu )^{1/2}}{(m_{0} - \mu  - |m_{12}|)^{1/2}} 
-
\frac{(m_{0} - \mu )^{1/2}}{(m_{0} - \mu  + |m_{12}|)^{1/2}} 
\Big\},
\nonumber 
\end{eqnarray}
where,  $\mu \equiv {\mu}_{n}, \, (n=1,2, \ldots, N)$, and, 
in the CPT invariant case: $m_{0} = H_{11}=H_{22}$, and 
${\Gamma}_{jk}, \,(j,k=1,2)$ can be defined as follows
\begin{equation}
{\Gamma}_{jk} f( \lambda ) =
\pi \sum_{n=1}^{N} g^{\ast}_{nj} (\lambda ) 
g_{nk}(\lambda ), \label{gamma-g}
\end{equation}
where, for simplicity, the weight function $f(\lambda )$ can be 
choosen analogously to (3.8) in \cite{chiu}. 

Now, following \cite{chiu,chiu1} one can identify 
${\Gamma}_{jk}, \, (j,k =1,2)$  with
those appearing in  the LOY theory (\ref{h-LOY-jk}),
$m_{0}$  can be  considered  as  kaon mass
\cite{chiu}, $m_{jk} \equiv H_{jk} \, (j,k =1,2)$, $\mu$ can
be treated as the mass of the decay products of  the
neutral  kaon \cite{chiu}. The additional assumption 
$|m_{12}| \ll (m_{0}- \mu )$ 
leads  to   the following estimation for 
$(h_{11}^{Imp}  -  h_{22}^{Imp})$:
\begin{equation}
h_{11}^{Imp}  -  h_{22}^{Imp}  
\simeq 
i \frac{ m_{21}{\Gamma}_{12} - m_{12}{\Gamma}_{21} }{4(m_{0} - \mu )}
\label{FL2}
\end{equation}
An equivalent form of this estimation is the following one:
\begin{equation}
h_{11}^{Imp}  -  h_{22}^{Imp} =  
\frac{- {\rm Re}( m_{12}) \; {\rm Im}({\Gamma}_{12}) + 
{\rm Im}(m_{12})\;{\rm Re}({\Gamma}_{12}) }{2(m_{0} - \mu )}. 
\label{FL3} 
\end{equation}
Real properties of neutral K--complex enable us to 
conclude that the
contribution of ${\rm Im}\,({\Gamma}_{12})$ in the numerator of 
(\ref{FL3})  is neglegibly small in comparison with the contribution 
of ${\rm Re}\,({\Gamma}_{12})$ in the considered case of neutral 
K--mesons  \cite{dafne}. Finally, taking into account that 
$\; 2 {\rm Re}\, ({\Gamma}_{12}) \simeq ({\gamma}_{s} - {\gamma}_{l})$ 
\cite{dafne},  the estimation (\ref{FL3}) takes the following 
form:
\begin{eqnarray}
h_{11}^{Imp}  -  h_{22}^{Imp}
& = &
{\rm Im}(m_{12})\; \frac{ {\gamma}_{s} - {\gamma}_{l} }
{4(m_{0} - \mu )}  \nonumber  \\
& \approx &
{\rm Im}(m_{12}) \;\frac{ {\gamma}_{s} }{4(m_{0} - \mu )}.
\label{FL4}
\end{eqnarray}
For the neutral K--system, to evaluate $(h_{11}^{Imp}  -  
h_{22}^{Imp})$ one can take 
${\tau}_{s} \simeq 0,89 \times 10^{-10} {\rm sec}$ \cite{data}. Hence
${\gamma}_{s} = \frac{\hbar}{{\tau}_{s}} \sim 7,4 \times 10^{-12} 
{\rm MeV}$ and (following \cite{chiu} ) $(m_{0} - \mu ) = m_{K} - 
2m_{\pi} \sim 200$ MeV. Thus 
\begin{equation}
(h_{11}^{Imp}  -  h_{22}^{Imp})
\sim  0,93 \times 10^{-14} {\rm 
Im}\,(m_{12})\equiv 0,93 \times 10^{-14} {\rm Im}\,(H_{12}).
\label{FL5}
\end{equation}


\begin{thebibliography}{10} \vspace*{-10pt}
\bibitem{LOY1} T. D. Lee, R. Oehme  and  C.  N.  Yang,  Phys.  Rev.,
{\bf 106}, (1957) 340.   \vspace*{-10pt}
\bibitem{LOY2} T. D. Lee and C. S.  Wu,  Annual  Review  of  Nuclear
Science, {\bf 16}, (1966) 471. \vspace*{-10pt}
\bibitem{Gaillard} Ed.:  M.  K.   Gaillard   and   M. Nikolic, Weak
Interactions, (INPN et de Physique des Particules,  Paris,  1977);
Chapt. 5, Appendix A. \vspace*{-10pt}
\bibitem{Bil-Kab} S. M. Bilenkij, Particles and nucleus, vol. 1, No 1
(Dubna 1970), p. 227 [in Russian]. P.  K.  Kabir,  The  CP-puzzle,
Academic Press,  New York 1968.  \vspace*{-10pt}
\bibitem{4} J. W. Cronin, Rev. Mod. Phys. {\bf 53}, (1981) 373.
J. W. Cronin, Acta Phys. Polon., {\bf B15}, (1984) 419.
V. V. Barmin, et al., Nucl. Phys. {\bf B247}, (1984) 293.
L. Lavoura, Ann. Phys. (NY), {\bf 207}, (1991) 428.
C. Buchanan, et al., Phys. Rev. {\bf D45}, (1992) 4088.
C. O. Dib, and R. D. Peccei, Phys. Rev.,{\bf D46}, (1992)
2265.
R. D. Peccei, CP and  CPT  Violation:  Status  and  Prospects,
Preprint  UCLA/93/TEP/19,  University  of  California,  June
1993.    \vspace*{-10pt}
\bibitem{5} E. D. Comins and P. H. Bucksbaum, Weak interactions of
Leptons and Quarks, (Cambridge University Press, 1983).
T. P. Cheng and L. F. Li, Gauge Theory of Elementary
Particle Physics, (Clarendon, Oxford 1984). \vspace*{-10pt}
\bibitem{dafne} L. Maiani, in "The Second Da$\Phi$ne Physics 
Handbook", vol. 1, Eds. L. Maiani, G. Pancheri and N. Paver, SIS 
--- Pubblicazioni, INFN  --- LNF, Frascati, 1995; pp. 3 
--- 26. \vspace*{-10pt}
\bibitem{leonid} L. A. Khalfin, Theory of Unstable 
Neutrino Mixing and the 17 keV Neutrino Problem, preprint 
HU--TFT-92--21, Helsinki, June 1992. \vspace*{-10pt}
\bibitem{baldo} M. Baldo--Ceolin, Neutron--antineutron oscillation 
experiments, Proceedings of the "International Conference of Unified 
Theories and Their Experimental Tests" --- Venice --- 16--18 March 
1982; 
Sensitive Search for Neutron--Antineutron Transitions at the Ill 
Reactor, AIP Conference Proceedings No 125 of the Fifth International 
Symposium on "Capture Gamma--Ray Spectroscopy and Related Topics" --- 
Edited by S. Raman,  Knoxville, September 1984. p.871. \vspace*{-10pt}
\bibitem{beyond} L. A. Khalfin,  The  theory   of
$K_{0}$,${\overline K}_{0}$  ($D_{0}$,${\overline D}_{0}$   and
$T_{0}$,${\overline T}_{0}$) mesons beyond the Weisskopf-Wigner
approximation  and  the
CP--problem, preprint LOMI P--4--80,  Leningrad,  February  
1980. \vspace*{-10pt}
\bibitem{leonid-fp} L. A. Khalfin, Foundations of Physics, {\bf
27} (1997) 1549. \vspace*{-10pt}
\bibitem{leonid1} L. A. Khalfin,  Preprints of the CPT, The University 
of Texas at  Austin: DOE-ER-40200-211,  February  1990  and
DOE-ER-40200-247, February 1991; (unpublished, cited in \cite{chiu}), 
and references one can find therein. \vspace*{-10pt}
\bibitem{chiu} C. B. Chiu and E. C. G. Sudarshan, Phys. Rev.
{\bf D 42} (1990) 3712. \vspace*{-10pt}
\bibitem{chiu1} E. C. G. Sudarshan, C. B. Chiu and
G. Bhamathi, Unstable Systems in Generalized Quantum Theory,
preprint DOE-40757-023 and CPP-93-23, University of Texas,
October 1993. \vspace*{-10pt}
\bibitem{PLB} K. Urbanowski, Phys. Lett. {\bf B 313},
(1993) 374.   \vspace*{-10pt}
\bibitem{is0} K. Urbanowski, Is  the  new  interpretation  of  some
standard CPT--violation parameters  necessary?,  Preprint  of  the
Pedagogical University, No WSP--IF 94--39, Zielona Gora,
May 1994;  CPT transformation properties of the exact effective 
Hamiltonian for neutral kaon complex, Preprint of the Pedagogical 
University, No WSP--IF 96--44, Zielona Gora, March 1996;
CPT transformation properties of the exact effective Hamiltonian 
for neutral kaon and similar complexes, Preprint of the Pedagogical 
University, No WSP--IF 97--50, Zielona Gora, July 1997
--  {\bf hep--ph/9803376}. \vspace*{-10pt}
\bibitem{is} K. Urbanowski, 
Int. J. Mod.  Phys.  {\bf  A 13}, (1998), 965.   \vspace*{-10pt}
\bibitem{why}K. Urbanowski, Why the LOY model 
cannot be used for designing CPT--invariance tests in neutral kaons 
systems? On CPT--noninvariance of systems with exponentially decaying 
particles, preprint of the Pedagogical University No WSP--IF 94--38, 
Zielona Gora, May 1994.\vspace*{-10pt}
\bibitem{tsai} T. Mochizuki, N. Hashimoto, A. Shinobori, S. Y. Tsai, 
Remarks on Theoretical Frameworks Describing The Neutral Kaon System, 
preprint Of The Nihon University, No NUP--A--97--14, Tokyo, June 
1997. \vspace*{-10pt}
\bibitem{ww} V. F. Weisskopf and E. T. Wigner, Z. Phys. 
{\bf 63} (1930) 54; \vspace*{-10pt} {\bf 65} (1930) 18. 
\bibitem{kabir} P. K. Kabir and A. Pilaftsis, Phys. Rev. {\bf A 53}, 
(1966) 66. \vspace*{-10pt}
\bibitem{cpt} W. Pauli, in: "Niels Bohr and the Developmnet of
Physics". ed. W. Pauli (pergamon Press, London, 1955), pp. 30 ---
51. G. Luders, Ann. Phys. (NY) {\bf 2} (1957) 1. . R. Jost, Helv.
Phys. Acta {\bf 30} (1957) 409.R.F. Streater and A. S. Wightman,
"CPT, Spin, Statistics and All That" (Benjamin, New York, 1964).
N. N. Bogolubov, A. A. Logunov and I. T. Todorov, "Introduction
to Axiomatic Field Theory" (Benjamin, New York, 1975). \vspace*{-10pt}
\bibitem{messiah} A. Messiah, Quantum Mechanics, vol. 2, (Wiley,
New York 1966). \vspace*{-10pt}
\bibitem{bohm} A. Bohm, Quantum Mechanics: Foundations and
Applications, 2nd ed., (Springer, New York 1986). \vspace*{-10pt}
\bibitem{ur1} K. Urbanowski, Int. J. Mod.  Phys.  {\bf A 10}, (1995)
1151.     \vspace*{-10pt}
\bibitem{master} e.g., R. Zwanzig, Physica {\bf 30}, (1964), 1109. 
F. Haake, "Statistical Treatment of Open Systems by Generalized Master 
Equations", Springer Tracts in Modern Physics, Vol. 66, (Springer, 
Berlin, 1973). L. P. Horwitz and J. P. Marchand, Rocky Mount. J. 
Math. {\bf 1}, (1971), 225. E. B. Davis, "Quantum Theory of Open 
System", (Academic, London,  \vspace*{-10pt}  1976). 
\bibitem{K-R1} W. Krolikowski and J. Rzewuski, Bull. Acad. Polon.
Sci. {\bf 4} (1956) 19. \vspace*{-10pt}
\bibitem{K-R2} W. Krolikowski and J. Rzewuski, Nuovo. Cim.
{\bf B 25} (1975) 739 and refernces therein. \vspace*{-10pt}
\bibitem{K-R-ur} K. Urbanowski, Acta Phys. Polon. {\bf B 14} (1983)
485. \vspace*{-10pt}
\bibitem{ur2} K. Urbanowski, Int. J. Mod.  Phys.  {\bf  A 8},  
(1993)3721. \vspace*{-10pt}
\bibitem{pra} K. Urbanowski, Phys. Rev. {\bf A 50}, (1994) 
2847. \vspace*{-10pt}
\bibitem{Cohen} E. Cohen and L. P. Horwitz, 
Preprints No {\bf hep--ph/9808030, hep--ph/9811332}. 
\vspace*{-10pt} 
\bibitem{fermilab} 
L. K. Gibbons et al., CP and CPT Symmetry Tests from the Two--pion
Decays of the Neutral Kaon with the Fermilab--E731 Detector,
preprints FERMILAB--Pub--95/392--E, EFI--95--76; Phys.Rev. 
{\bf D55}, (1997), 6625. \vspace*{-10pt}
\bibitem{cpt-qm}
J. Ellis, J.L. Lopez, N.E. Mavratos and D.V. Nanopulous, Phys. Rev. 
{\bf D53}, (1996), 3846;
J. Ellis,  N.E. Mavratos and D.V. Nanopulous, Phys. Lett.,
{\bf B 293},(1992), 142;
V.A. Kostelecky, Phys. rev. Lett., {\bf 80}, (1998), 1818.
\vspace*{-10pt}
\bibitem{5th-f}
E. Fischbach, C. Talmadge, Nature, {\bf 356}, (1992), 207; 
E. Fischbach, D. Sudarsky, A. Szafer, C. Talmadge and
S. H. Aronson,Phys. Rev. Lett., {\bf 56}, (1986), 3; 
D. Sudarsky, E. Fischbach, C. Talmadge, A. H. Aronson and H.--Y.
Cheng, Annals of Physics, {\bf 207}, (1991), 103; 
E. Fischbach, C. Talmadge, Preprint No {\bf hep--ph/9606249}.
\vspace*{-10pt}
\bibitem{sr-eqp}
T. Haymbye, R. B. Mann, U. Sarkar, Phys. Lett., {\bf B421}, (1998),
105; 
T. Haymbye, R. B. Mann, U. Sarkar, Phys. Rev. {\bf D58}, (1998),
art. no 025003; 
C. Alvarez, R. B. Mann, Phys. Rev., {\bf D55}, (1997), 1732;
R. J. Hughes, Phys. Rev., {\bf 46}, (1992), R2283;
I. R. Kenyon, Phys. Lett., {\bf B 237}, (1990), 274. \vspace*{-10pt}
\bibitem{data}
Review of Particle Physics, Phys. Rev. {\bf D54}, (1996), No 1, 
Part 1.
\end{thebibliography}
\end{document}